\documentclass[
floats,floatfix,
nofootinbib,
prd,aps,superscriptaddress,tightenlines]{revtex4}
\usepackage{graphicx}
\usepackage{bm}
\usepackage{pstricks}

\def\OMIT#1{{}}

\newcommand{\bea}{\begin{eqnarray}}
\newcommand{\eea}{\end{eqnarray}}
\newcommand{\beq}{\begin{equation}}
\newcommand{\eeq}{\end{equation}}
\newcommand{\bay}{\begin{array}}
\newcommand{\eay}{\end{array}}
\newcommand{\vslash}{\mbox{$\not{\hspace{-1.03mm}v}$}}        
\newcommand{\Dslash}{\mbox{$\not{\hspace{-1.03mm}D}$}}        
\newcommand{\qslash}{\mbox{$\not{\hspace{-0.8mm}q}$}}
\newcommand{\Dleft}{\overleftarrow D}
\newcommand{\Dright}{\overrightarrow D}

\begin{document}
\begin{flushright}
CTP-MIT-3490
\end{flushright}

\title{Exclusive rare $B\to K^* \ell^+\ell^-$ decays at low recoil:\\ 
controlling the long-distance effects} 

\author{Benjam\'\i{}n Grinstein}
\affiliation{Department of Physics, UCSD, 9500 Gilman Drive, La Jolla, CA 92093}

\author{Dan Pirjol}
\affiliation{Center for Theoretical Physics, Massachusetts Institute of 
Technology, Cambridge, MA 02139}

\date{\today }

\begin{abstract}
\vspace{1.0cm}

We present a model-independent description of the exclusive rare
decays $\bar B\to K^* e^+e^-$ in the low recoil region (large lepton
invariant mass $q^2\sim m_b^2$).  In this region the long-distance
effects from quark loops can be computed with the help of an operator
product expansion in $1/Q$, with $Q=\{m_b, \sqrt{q^2}\}$.
Nonperturbative effects up to and including terms suppressed by $\Lambda/Q$
and $m_c^2/m_b^2$ relative to the short-distance amplitude can be
included in a model-independent way.  Based on these results, we
propose an improved method for determining the CKM matrix element
$|V_{ub}|$ from a combination of rare and semileptonic $B$ and $D$
decays near the zero recoil point. The residual theoretical
uncertainty from long distance effects in this $|V_{ub}|$
determination comes from terms in the OPE of order $\alpha_s(Q)\Lambda/m_b,
\alpha_s^2(Q), m_c^4/m_b^4$ and duality violations, 
and is estimated to be below $10\%$.
\end{abstract}

\maketitle

\section{Introduction}

Radiative $B$ decays are important sources of information about the
weak couplings of heavy quarks. Experiments at the B factories have
measured precisely the branching ratios of the exclusive rare radiative 
$b\to s\gamma$ and semileptonic $b\to u e\nu$ decays, and decay spectra 
are beginning to be probed.
In addition to offering ways of extracting the CKM matrix
elements $V_{ub}$ and $V_{td}$, these processes hold good promise for the
detection of new physics effects (see e.g. \cite{ABHH}).

In contrast to the inclusive heavy hadron decays which can be reliably
described using the heavy mass expansion, the corresponding
heavy-light exclusive decays are comparatively less well
understood. The theoretical ignorance of the strong interaction
effects in these decays is parameterized in terms of unknown heavy to
light $B\to M$ form factors. Although lattice \cite{lattice} and QCD sum 
rules \cite{QCDSR} have
made significant progress in computing these form factors, they
are still beset with large errors and limitations. 

In the low recoil region, heavy quark symmetry has been used to relate
some of the $B\to M$ form factors \cite{IsWi,BuDo}. In Refs.~\cite{GrPi1,GrPi2} we
showed that the leading corrections to these symmetry relations when
$m_b\neq\infty$ do not involve any non-local contributions, that is, they are
characterized solely in terms of matrix elements of local
operators. Here we show that the cancellations of non-local terms,
which appear as a remarkable accident in the heavy quark effective
theory, are easily understood by deriving the form factors relations
directly from QCD at finite $m_b$. 

For the case of $b\to s e^+ e^-$ decays there is an additional source of theoretical
uncertainty due to long distance effects involving  the weak
nonleptonic Hamiltonian and the quarks' electromagnetic current.
In $B\to K^* e^+ e^-$, these effects are numerically significant
for a dilepton invariant mass close to the $c\bar c$
resonance region $q^2=(p_{e^+} + p_{e^-})^2 \sim 10$~GeV$^2$.
Usually these effects are computed using the parton model 
\cite{Heff,BuMu}, or vector meson dominance, by assuming saturation
with a few low lying resonances $\psi_n$ and using the factorization 
approximation for the nonleptonic decay amplitudes $B\to K^* \psi_n$ 
\cite{VMD,LSW,ABHH}.
Such a procedure is necessarily model dependent, and its effect on the 
$|V_{ub}|$ determination has been estimated at $\sim 10\%$.
Although in principle the validity of the approximations made can be 
tested {\em aposteriori}
by measuring other predicted observables, such as the shape of the $q^2$ spectrum
or angular distributions, it is clearly
desirable to have a more reliable computation of these effects.

The object of this paper is to show that, near the zero recoil point
$q^2 \sim q^2_{\rm max}=(m_B-m_{K^*})^2$, these long distance
contributions to $B\to K^* e^+ e^-$ can be computed as a short-distance
effect using simultaneous heavy quark and operator product expansions
in $1/Q$, with $Q = \{ m_b, \sqrt{q^2} \}$.
We use this expansion to develop a power counting scheme for the
long-distance amplitude, and classify the various contributions in
terms of matrix elements of operators. The leading term in the
expansion is calculated in terms of the form factors that were
necessary to parametrize the local, leading contribution to the decay
amplitude. Moreover, the first correction, of order $\Lambda/Q$, is given in
terms of the same operators introduced in Ref.~\cite{GrPi1} to parameterize 
the leading
order corrections to the heavy quark symmetry relations between form
factors, and is suppressed further by a factor of $\alpha_s(m_b)$. The
largest second order correction, of order $z = m_c^2/m_b^2$, is also
calculable in terms of the leading form factors. Hence, our method for
computing the long distance contributions introduces no new model dependencies to
good accuracy.  The terms we neglect are suppressed by $m_c^4/m_b^4$
and $\Lambda^2/m_b^2$ relative to the short-distance amplitude, and are
expected to introduce an uncertainty in $|V_{ub}|$ of about $1-2\%$.

A model-independent determination of $|V_{ub}|$ has been proposed
using semileptonic and rare B and D decays in the low recoil
kinematic region \cite{IsWi,SaYa,LiWi,LSW}.  This method uses
heavy quark symmetry to relate the semileptonic and rare radiative B form factors. 
More specifically, this method requires the rare and
semileptonic modes $\bar B\to K^* e^+ e^-$,  $\bar B\to \rho e\nu$, $\bar D\to
K^* e\nu$ and $\bar D\to \rho e\nu$. The main observation is that, neglecting
the long distance contribution to the radiative decay, the double
ratio $[\Gamma(\bar B\to K^* e^+ e^-)/ \Gamma(\bar B\to \rho e\nu)]/[\Gamma(\bar D\to K^*
e\nu)/ \Gamma(\bar D\to \rho e\nu)]$ is calculable since it is protected by both
heavy quark and $SU(3)$-flavor symmetries \cite{Grin}. We extend this result to
include the long distance contributions which, as explained above, are
calculable in terms of the same form factors in the endpoint
region. 

The modes required for this determination are beginning to be probed experimentally. 
The branching ratios of the rare decays $B\to K^{(*)}\ell^+\ell^-$ have
been measured by both the BABAR \cite{Babar} and BELLE \cite{Belle} (with
$\ell = e,\mu$)
collaborations
\begin{eqnarray}
{\cal B}(B\to K^* \ell^+ \ell^-) = 
\left\{
\begin{array}{ll}
(0.88^{+0.33}_{-0.29}\pm 0.10) \times 10^{-6} &  \mbox{(BABAR)} \\
(11.5 ^{+2.6}_{-2.4}\pm 0.8 \pm 0.2)\times 10^{-7} & \mbox{(BELLE)} \\
\end{array}
\right.
\nonumber
\end{eqnarray}
and
\begin{eqnarray}
{\cal B}(B\to K \ell^+ \ell^-) = 
\left\{
\begin{array}{ll}
(0.65^{+0.14}_{-0.13}\pm 0.04) \times 10^{-6} & \mbox{(BABAR)}\\
(4.8 ^{+1.0}_{-0.9}\pm 0.3 \pm 0.1)\times 10^{-7} & \mbox{(BELLE)}\,.
\nonumber
\end{array}
\right.
\end{eqnarray}
This suggests that a determination of $|V_{ub}|$
using these decays might become feasible in a not too distant future.

The paper is organized as follows. In Sec.~II we construct the operator
product expansion (OPE) formalism for the long-distance contribution to
exclusive $B\to K^* e^+ e^-$ decay in the low recoil region 
$q^2 \sim q_{\rm max}^2$. This is formulated as an expansion in $1/Q$,
with $Q = \{ m_b, \sqrt{q^2} \}$. The coefficients of the operators 
in the OPE are determined by matching at the scale $Q$, which is discussed in
some detail in Sec.~III. In Sec.~IV we present the evaluation of the hadronic
matrix elements of the operators appearing in the OPE, and explicit results
for the $|V_{ub}|$ determination are presented in Sec.~V. An Appendix contains
a simplified derivation of the improved form factor symmetry relations at low
recoil.

\section{Operator product expansion}
\label{sec2}

The effective Hamiltonian mediating the rare decays $b\to s e^+ e^-$ is \cite{Heff}
\bea\label{Hw}
{\cal H}_{\rm eff} = -\frac{G_F}{\sqrt2} V_{tb} V_{ts}^* \sum_{i=1}^{10} C_i(\mu)
Q_i(\mu)\,,
\eea
where the operators $Q_i$ can be chosen as
\begin{eqnarray}\label{Qi}
Q_1 &=& (\bar s_\alpha c_\beta)_{V-A} (\bar c_\beta b_\alpha)_{V-A}\\
Q_2 &=& (\bar s c)_{V-A} (\bar c b)_{V-A}\nonumber\\
Q_3 &=& (\bar s b)_{V-A} \sum_q (\bar q q)_{V-A}\nonumber\\
Q_4 &=& (\bar s_\alpha b_\beta)_{V-A} \sum_q (\bar q_\beta q_\alpha)_{V-A}\nonumber\\
Q_5 &=& (\bar s b)_{V-A} \sum_q (\bar q q)_{V+A}\nonumber\\
Q_6 &=& (\bar s_\alpha b_\beta)_{V-A} \sum_q (\bar q_\beta q_\alpha)_{V+A}\nonumber\\
Q_7 &=& \frac{e}{8\pi^2} m_b \bar s_\alpha \sigma_{\mu\nu} (1+\gamma_5)
b_\alpha F_{\mu\nu}\nonumber\\
Q_8 &=& \frac{g}{8\pi^2} m_b \bar s_\alpha \sigma_{\mu\nu} (1+\gamma_5)
T^a_{\alpha\beta} b_\beta G^a_{\mu\nu}\nonumber\\
Q_9 &=& \frac{e^2}{8\pi^2}(\bar sb)_{V-A} (\bar e e)_{V}\nonumber\\
Q_{10} &=& \frac{e^2}{8\pi^2}(\bar sb)_{V-A} (\bar e e)_{A}\,.\nonumber
\end{eqnarray}
We denoted here $(\bar q q)_{V\pm A} = \bar q \gamma_\mu (1\pm \gamma_5) q$.
The contributions of the operators $Q_{7,9,10}$ are factorizable and can be 
directly expressed through form factors, while the remaining
operators $Q_{1-6}$ contribute through nonlocal matrix elements with the
quarks' electromagnetic coupling $j_{\rm e.m.}^\mu = \sum_q Q_q \bar q\gamma^\mu q$
as
\bea\label{AVAdef}
A(\bar B\to K^* e^+e^-) = \frac{G_F}{\sqrt2} V_{tb} V^*_{ts} \frac{\alpha}{2\pi}
\left\{ (\bar e\gamma^\mu e) A^{(V)}_\mu + 
(\bar e\gamma^\mu\gamma_5 e) A^{(A)}_\mu \right\}\,.
\eea
The two hadronic amplitudes $A_\mu^{(V,A)}$ are given explicitly by
\bea\label{AV}
A^{(V)}_\mu &=& -C_7(\mu) \frac{2m_b}{q^2} \langle K^*(k,\eta)|\bar si\sigma_{\mu\nu}
q^\nu (1+\gamma_5) b|\bar B(v)\rangle \\
& &+ C_9(\mu) \langle K^*(k,\eta)|\bar s \gamma_\mu (1-\gamma_5) b|\bar B(v)\rangle
- 8 \pi^2 \frac{1}{q^2} \sum_{i=1}^6  C_i(\mu) {\cal T}^{(i)}_\mu(q^2, \mu)\nonumber\\
\label{AA}
A^{(A)}_\mu &=& C_{10}(\mu)
\langle K^*(k,\eta)|\bar s \gamma_\mu (1-\gamma_5) b|\bar B(v)\rangle \,.
\eea
where we introduced the nonlocal matrix element parameterizing the
long-distance amplitude
\bea\label{Ti}
{\cal T}_i^\mu(q^2) = i \int d^4x e^{iq\cdot x} \langle K^*(k,\eta)|T
Q_i(0)\,, j_{\rm e.m.}^\mu(x) |\bar B(v)\rangle\,.
\eea
The conservation of the electromagnetic current implies in the usual
way the Ward identity (see e.g. \cite{GP0,GNR}) for the long-distance amplitude
\begin{eqnarray}\label{ward}
q^\mu {\cal T}_i^\mu(q^2) = 0\,.
\end{eqnarray}

Our problem is to compute ${\cal T}_i^\mu(q^2)$ in the low recoil
region, corresponding to $q^2 \sim m_b^2$. Consider the amplitude ${\cal
T}_i(q^2)$ as a function of the complex variable
$q^2$. This is an analytic function everywhere in the complex $q^2$
plane, except for poles and cuts corresponding to states with the
quantum numbers of the photon $J^{PC} = 1^{--}$. The region kinematically
accessible
in $B\to K^* e^+ e^-$ is the segment on the real axis $q^2 = [0,
q^2_{\rm max} = (m_B-m_V)^2]$.

This is very similar to $e^+ e^- \to $ hadrons, which is related by
unitarity to the correlator of two electromagnetic currents
$\Pi^{\mu\nu}(q^2) =\Pi(q^2)(q^\mu q^\nu-q^2 g_{\mu\nu}) = i\int d^4 x e^{iq\cdot x} \langle 0|T
j^\mu(0), j^\nu(x)|0\rangle $.  For this case, it is well known that at large
time-like $q^2$, both the dispersive and imaginary parts of the
correlator $\Pi(q^2)$ can be computed in perturbation theory. This is
the statement of local duality\cite{Bloom:1970xb}, which
is expected to hold up to power corrections in
$1/Q$\cite{PQW,Shifman:2000jv}.
In contrast to $e^+ e^-\to$ hadrons, the external states appearing in the definition
of ${\cal T}_i(q^2)$ are strongly interacting. For this reason, a closer analogy is to
the computation of the inclusive semileptonic width of $B$ hadrons 
using the OPE and heavy quark expansion\cite{heavyOPE}. 

The zero recoil
point in $B\to K^* e^+ e^-$  corresponds to a dilepton invariant mass 
$q^2_{\rm max}=(m_B - m_{K^*})^2 = 19.2$ GeV$^2$ and 
is sufficiently far away from the threshold of the
resonance region connected with $c\bar c$ states $q^2 \sim 10$ GeV$^2$.
Therefore duality can be expected to work reasonably well. 
There are, in addition, effects from thresholds of other  $J^{PC} =
1^{--}$ states, like the $\rho$ and the $\Upsilon$. These effects are smaller
because they either enter through the operators $Q_3$--$Q_6$, which
have small Wilson coefficients, or through $Q_1^u=(\bar s_\alpha u_\beta) 
(\bar u_\beta b_\alpha)$ and $Q_2^u=(\bar s u) (\bar u b)$ through
CKM suppressed loops $\sim V_{ub} V_{us}^*$. The effects of light states, like the $\rho$-meson, are
under better control since the associated resonance regions are
even lower than for $c\bar c$. Heavier states, like the $\Upsilon$,  lie above
$q^2_{\rm max}$. These too are under better control since duality sets in
much faster from below resonance than from above, as evidenced by empirical
observation, as in the example of $e^+ e^- \to $ hadrons.

In analogy with the OPE for the inclusive B decays, we propose to
expand the amplitudes ${\cal T}_i(q^2)$ in an operator product expansion in
the large scale $Q = \{m_b, \sqrt{q^2}\}$
\bea\label{OPE}
{\cal T}_i^\mu(q^2) = \sum_{k \geq -2}\sum_{j}
C^{(k)}_{i,j}(q^2/m_b^2, \mu) \langle {\cal O}^{(k)\mu}_j(\mu) \rangle
\eea
where the contribution of the operator ${\cal O}^{(k)}_j$ scales like
$1/Q^k$. 
The operators appearing on the right-hand side are constructed
using the HQET bottom quark field $h_v$, and they can contain
explicit factors of the velocity $v$ and the dilepton momentum
$q$. Their matrix elements must satisfy the Ward identity Eq.~(\ref{ward})
for all possible external states, which has therefore to be satisfied
at operator level. In addition, they must transform 
in the same way as 
${\cal T}_i^\mu$ under the chiral $SU_L(3) \times SU_R(3)$ 
group, up to factors of the light quark masses which can flip
chirality.

Our analysis will be valid in the small recoil region, where the 
light meson kinetic energy is small $E_V - m_V \sim \Lambda$.
Expressed in terms of the dilepton invariant mass $q^2$ this
translates into the range $(m_B-m_V)^2 - q^2 \leq 2m_B \Lambda$.
In the particular case of $B\to K^* e^+ e^-$ this region extends about 
5 GeV$^2$ below the maximal
value $q_{\rm max}^2 = (m_B - m_{K^*})^2 = 19.2$ GeV$^2$.

Each term in the OPE Eq.~(\ref{OPE}) must have mass 
dimension 5. The leading contributions come
from operators whose matrix elements scale like $Q^2$ 
\bea\label{LO1}
{\cal O}^{(-2)}_1 &=& \bar s_L [q^2 \gamma_\mu - q^\mu \qslash] h_{vL}\\
\label{LO2}
{\cal O}^{(-2)}_2 &=& 
im_b \bar s_L \sigma_{\mu\nu} q^\nu  h_{vR} \,.
\end{eqnarray}
Another allowed operator $(q^2 v_\mu - q_\mu v\cdot q)(\bar s_L
h_{vR})$ can be shown in fact to scale like $Q\Lambda$ after using 
Eq.~(\ref{q}), and is included below as ${\cal O}_3^{(-1)}$ (see Eq.~(\ref{NLO3})).
These operators are written in terms of chiral 
fields $q_{L,R} = P_{L,R} q$, with $P_{L,R} = \frac12 (1\mp \gamma_5)$.
In the chiral limit only $s_L$ can appear, and
the right-handed field $s_R$ requires an explicit factor of $m_s$.

In general, the dilepton momentum $q^\alpha$ can be rewritten as a
constant part plus a total derivative acting on the current
\begin{eqnarray}\label{q}
q^\alpha (\bar s\Gamma h_v) = (m_b v^\alpha + i\partial^\alpha)
(\bar s\Gamma h_v)\,,
\end{eqnarray}
where the two terms on the right-hand side scale like $m_b$ and $\Lambda$,
respectively. For this reason, using $q^\alpha$ in the definition of the
operators gives them a non-homogeneous scaling in $1/m_b$. This is not
a problem in the power counting scheme adopted here, which counts $m_b$
and $Q$ as being comparable. We will keep $q^\alpha$ explicit in the
leading operators Eq.~(\ref{LO1}), (\ref{LO2}), which
we would like to write in a form as close as possible to the 
short-distance operators.
On the other hand, we expand in $1/m_b$ in the sub-leading operators below,
and keep only the leading term in Eq.~(\ref{q}).

\begin{figure}[t!]
\begin{center}
  \hspace{-0.5cm}
  \includegraphics[width=2in]{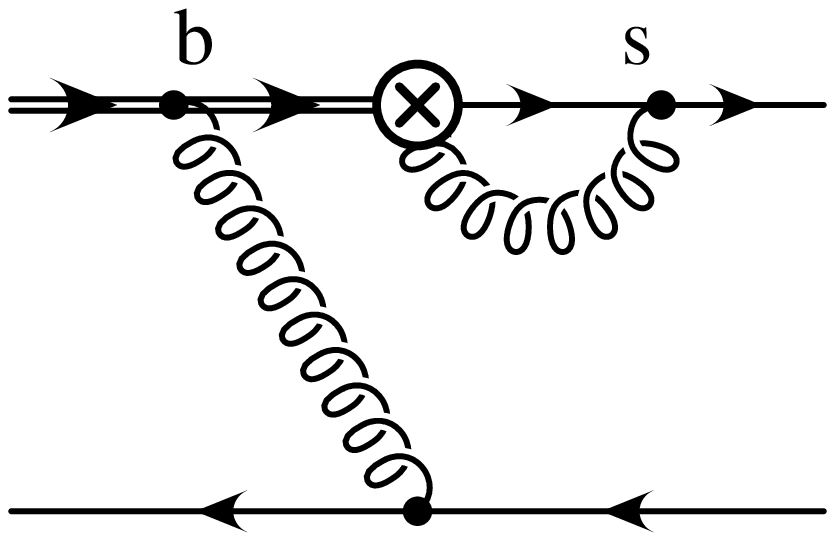}
\hspace{1cm}
  \includegraphics[width=2in]{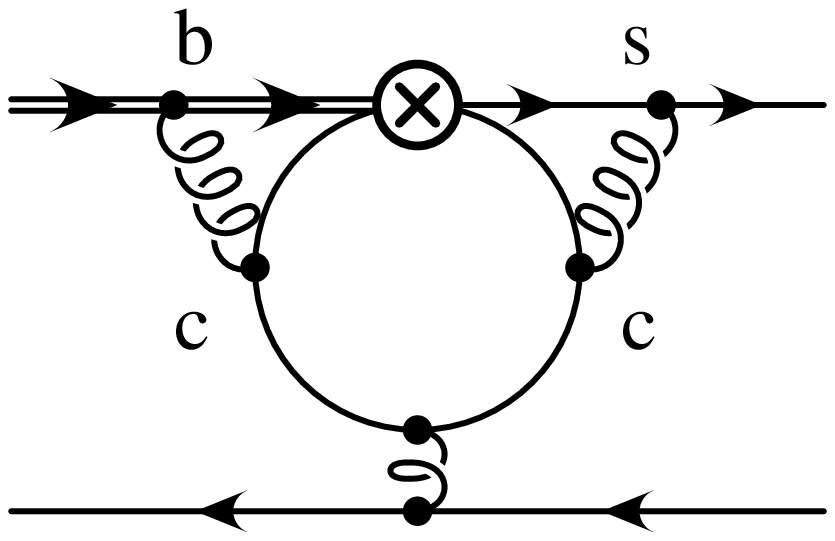}\\
\vspace{0.4cm}
\hspace{0.2cm} (a) \hspace{5.5cm} (b) \hspace{5cm} 
\end{center}
{\caption{
Contributions to the $B\to K^* \ell^+ \ell^-$ amplitude near the zero
recoil point coming from different operators in the OPE Eq.~(\ref{OPE}).
In (a) the circled cross denotes one of the operators ${\cal O}^{(-1,0)}$
of the form $\bar q \Gamma iD_\mu h_v$ or $\bar q \Gamma gG_{\mu\nu} h_v$, 
and in (b) it denotes one of the 4-quark operators $(\bar q h_v)(\bar c c)$.
The contributions in (a) are suppressed relative to the short-distance amplitude
by $\Lambda/Q$ (for ${\cal O}^{(-1)}$),
$\Lambda^2/Q^2$ (for ${\cal O}^{(0)}$), and those in (b) by $m_c^4/Q^4$.
}}
\end{figure}

Next we include operators whose matrix elements scale like $Q\Lambda$.
They are dimension-4 operators of the form $\bar q \Gamma iD_\mu h_v$.
A complete set of operators which satisfies the condition (\ref{ward})
and which do not vanish by the equations of motion can be chosen as
\bea\label{NLO1}
{\cal O}^{(-1)}_1 &=& m_b \bar s_L 
[i\Dleft_\mu - v_\mu (v\cdot i\Dleft)] h_{vR}\\
\label{NLO2}
{\cal O}^{(-1)}_2 &=&  m_b (v\cdot i\partial) \bar s_L 
[\gamma_\mu - v_\mu \vslash ] h_{vL}\\
\label{NLO3}
{\cal O}^{(-1)}_3 &=&   m_b [ i\partial_\mu - v_\mu (v\cdot i\partial) ]
(\bar s_L h_{vR} ) \\
\label{NLO4}
{\cal O}^{(-1)}_4 &=&   m_b i\partial_\nu (\bar s_L 
[\gamma_\mu - v_\mu \vslash ] \gamma^\nu h_{vR})\\
\label{NLO5}
{\cal O}^{(-1)}_5 &=&  m_b m_s \bar s_R (\gamma_\mu - v_\mu \vslash) h_{vR}\,.
\end{eqnarray}
The operator ${\cal O}^{(-1)}_5$ describes effects where one chirality flip
occurs on the light quark side. Its matrix element scales like
$Q m_s$.

There are no contributions scaling like $Q m_c$, since the dependence on
the charm quark mass  must contain only even powers of $m_c$. The leading contributions
containing $m_c$ scale like $m_c^2$ and come from operators similar to 
(\ref{LO1}) and (\ref{LO2}). We will define them as
\bea\label{NNLO1}
{\cal O}^{(0)}_1 &=& m_c^2 \bar s_L[\gamma_\mu - q_\mu \qslash/q^2 ] h_{vL}\\
\label{NNLO2}
{\cal O}^{(0)}_2 &=& 
im_b \frac{m_c^2}{q^2} \bar s_L \sigma_{\mu\nu} q^\nu  h_{vR} \,.
\eea

There are many operators whose matrix elements scale like $\Lambda^2$; generally,
they are of the form ${\cal O}^{(0)}_{3,\dots} = \bar q \Gamma (iD_\mu)(iD_\nu) h_v$ or contain one factor of
the gluon tensor field strength $\bar q \Gamma gG_{\mu\nu} h_v$. The latter operators 
can appear at $O(\alpha_s^0)$ in matching from graphs with $q\bar q$ quark loops
as shown in Fig.~2(c), and can contribute to the $B\to K^* \ell^+\ell^-$ amplitude
through the graph in Fig.~1(a). 

Another class of operators appearing in the OPE describes effects of propagating
charm quarks (see Fig.~1(b)), and have the form
\begin{eqnarray}\label{cpengs}
{\cal O}^{(2)} = \frac{1}{Q^2} (\bar s \Gamma h_v)(\bar c \Gamma_c iD_\mu c)\,.
\end{eqnarray}
The explicit form of these operators will be given in the next section, where it 
is shown that their
contributions are further suppressed by $m_c^4/Q^4$ relative to the short-distance
amplitude.  

To sum up the discussion of this section, we argued that the long-distance
effects to $b\to s \ell^+ \ell^-$ decays in the zero recoil region come from
well-separated scales satisfying the hierarchy $m_b\sim Q > m_c > \Lambda$. 
These effects can be resolved using an OPE as shown in Eq.~(\ref{OPE}). The contributions
of the various operators in the OPE, relative to the dominant short-distance amplitude, 
are summarized in Table \ref{powercounting}, together with the order in matching 
(in $\alpha_s(Q)$) at which they start contributing. 

Some of the subleading operators appearing in the OPE give spectator type 
contributions to the exclusive $B\to K^* \ell^+\ell^-$ amplitude, as shown in
Fig.~1. For example, the $O(\Lambda Q)$ operators ${\cal O}_j^{(-1)}$ and 
$O(\Lambda^2)$ operators ${\cal O}_j^{(0)}$ can contribute through the
graphs in Fig.~1(a), and the charm operators of the type Eq.~(\ref{cpengs}) 
contribute as in Fig.~1(b). Such spectator type contributions were studied at 
lowest order in perturbation theory in \cite{LiWi} where they were shown to be suppressed 
at least by $\Lambda/Q$. The effective theory approach used here extends this proof to
all orders in $\alpha_s$, and shows that the suppression factor is 
$\alpha_s(Q) \Lambda/Q$
(for the contributions from ${\cal O}_j^{(-1)}$) and $\Lambda^2/Q^2$ (for contributions
coming from ${\cal O}_j^{(0)}$).

We comment briefly on an alternative approach used in Refs.~\cite{BuMu,LiWi}
where the charm quarks and the large scales $\sqrt{q^2}, m_b$ are integrated 
out simultaneously. Such an approach includes the charm mass effects to
all orders in $m_c^2/m_b^2$, but has the disadvantage of introducing 
potentially large power corrections $\sim \Lambda^2/m_c^2$. For this reason
we prefer to integrate out only the large scale $Q$ and leave the charm as 
a dynamical field in the OPE.

The main result of our
paper is that the contributions of leading order $O(1)$ and the power suppressed
terms $O(m_c^2/Q^2)$ to the long-distance amplitude depend only on
known form factors and thus can be included {\em without
introducing any new hadronic uncertainty}. The power suppressed terms 
of $O(\Lambda/Q)$ can be accounted for in terms of the form factors of
the two dimension-4 currents $\bar q iD_\mu (\gamma_5) h_v$.

\begin{table}[t!]
\begin{center}
\begin{tabular}{ccc}
\hline
Operator & Power counting & Order in matching \\
\hline\hline
${\cal O}_{1,2}^{(-2)}$  & 1                &  $\alpha_s^0(Q)$ \\
${\cal O}_{1-5}^{(-1)}$  & $\Lambda/Q$     & $\alpha_s(Q)$  \\
${\cal O}_{1,2}^{(0)}$   & $m_c^2/Q^2$     & $\alpha_s^0(Q)$  \\
${\cal O}_{j>3}^{(0)}$    & $\Lambda^2/Q^2$ &  $\alpha_s^0(Q)$  \\
${\cal O}_i^{(2)}$       & $m_c^4/Q^4$     & $\alpha_s^0(Q)$ \\
\hline
\end{tabular}
\end{center}
{\caption{Contributions to the long-distance amplitude for
$b\to s \ell^+\ell^-$ coming from the different operators in the OPE
Eq.~(\ref{OPE}), together with the order in $\alpha_s(Q)$ at which
they appear in matching.}
\label{powercounting} }
\end{table}

In the next Section we compute the matching conditions for these operators at
lowest order in perturbation theory.

\section{Matching}
\label{sec3}

Typical lowest order diagrams contributing to the $T-$products ${\cal T}_i^\mu(q^2)$
in QCD are shown in Fig.~1. The matching conditions for the operators appearing
in the OPE Eq.~(\ref{OPE}) are found by computing these graphs and expanding them in
powers in $1/Q$. At lowest order in $\alpha_s(Q)$ the graph in
Fig.~1(a) will match onto ${\cal O}^{(-2)}_j$, but not onto the $O(\Lambda Q)$ operators
${\cal O}^{(-1)}_j$. These operators appear first at $O(\alpha_s(Q))$ from 
graphs containing one additional gluon as shown in Fig.~1(b).

\begin{figure}[t!]
\begin{center}
  \hspace{-0.5cm}
  \includegraphics[width=1.5in]{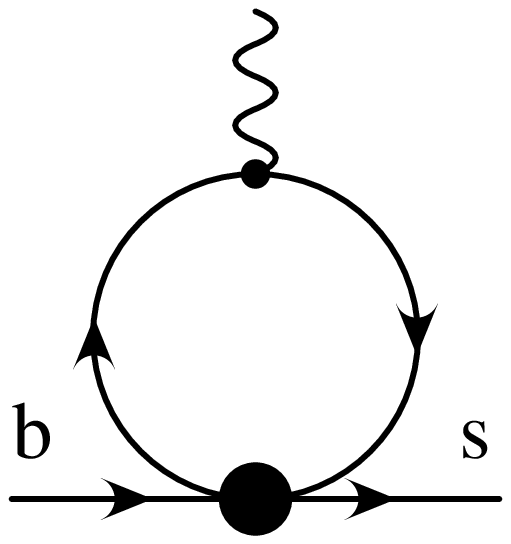}\hspace{1cm}
\includegraphics[width=2.0in]{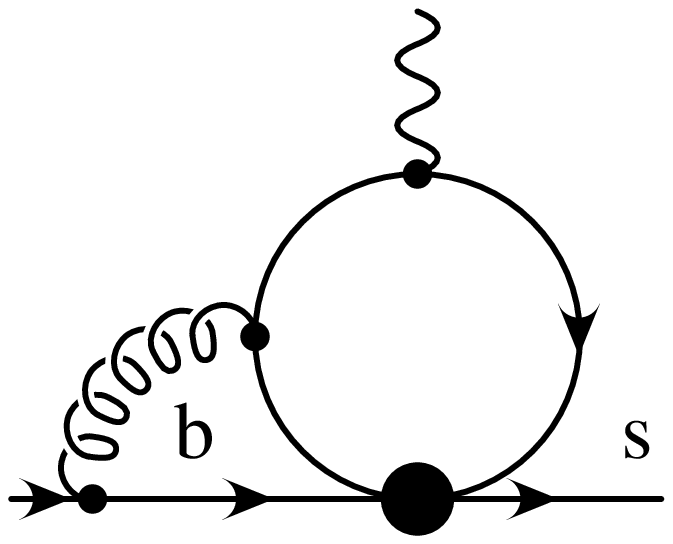}\hspace{1cm}
\includegraphics[width=1.5in]{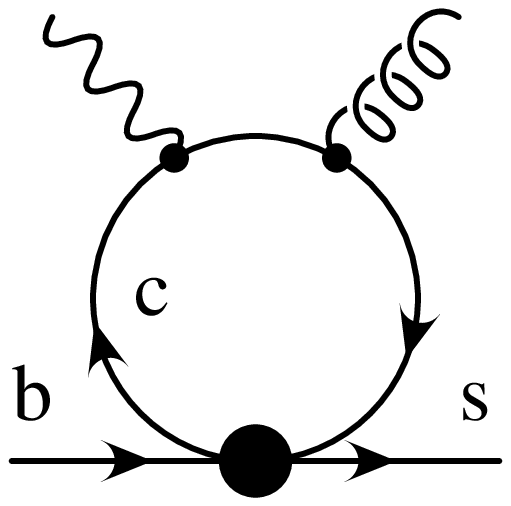} \\
\hspace{0.5cm} (a) \hspace{5cm} (b) \hspace{5cm} (c)
\end{center}
{\caption{
Graphs in QCD contributing to the matching onto $\bar s\Gamma h_v$ operators 
(a), $\bar s\Gamma iD_\mu h_v$ (b) and $\bar s gG^{\mu\nu} \Gamma_\nu h_v$ operators (c). 
The filled circle  denotes the insertion of $Q_{1-6}$. In (c) the wavy line is the 
virtual photon $\gamma^*$ and the curly line denotes a gluon.}}
\end{figure}

An explicit computation of the graph in Fig.~1(a) with one insertion of the 
operators $Q_{1-6}$  gives the following results for the matrix elements of the
$T-$products ${\cal T}_i^\mu(q^2)$ on free quark states \cite{Heff,BuMu}
(we use everywhere naive dimensional regularization (NDR) with an anticommuting
$\gamma_5$ matrix)
\bea\label{loop1}
\langle {\cal T}_1^\mu(q^2) \rangle &=& \frac{1}{2\pi^2}
\langle \bar s(q^2 \gamma^\mu - q^\mu \qslash)P_L b \rangle
\left\{ -\frac{2}{3\epsilon} + \frac23 + 4 G(m_c)\right\}\\
\label{loop2}\langle {\cal T}_2^\mu (q^2) \rangle &=& \frac{1}{2\pi^2}
\langle \bar s(q^2 \gamma^\mu - q^\mu \qslash)P_L b \rangle
\left\{ -\frac{2}{9\epsilon} + \frac29 + \frac{4}{3}G(m_c)\right\}\\
\label{loop3}\langle {\cal T}_3^\mu(q^2) \rangle &=& \frac{1}{2\pi^2}
\langle \bar s(q^2 \gamma^\mu - q^\mu \qslash)P_L b \rangle
\left\{ -\frac{1}{9\epsilon} - \frac{2}{9} + 4G(m_c) - \frac23 G(0) -
\frac83 G(m_b)\right\}\\
\label{loop4}\langle {\cal T}_4^\mu(q^2) \rangle &=& \frac{1}{2\pi^2}
\langle \bar s(q^2 \gamma^\mu - q^\mu \qslash)P_L b \rangle
\left\{ \frac{5}{9\epsilon} - \frac23 - 2G(0)
+ \frac43 G(m_c) - \frac83 G(m_b)\right\}\\
\label{loop5}\langle {\cal T}_5^\mu(q^2) \rangle &=& \frac{1}{2\pi^2}
\langle \bar s(q^2 \gamma^\mu - q^\mu \qslash)P_L b \rangle
\left\{ -\frac{1}{3\epsilon} + 4G(m_c) - 2G(m_b) \right\}
- \frac{1}{6\pi^2}
m_b \langle \bar s (\qslash \gamma_\mu - q_\mu) P_R b \rangle \\
\label{loop6}\langle {\cal T}_6^\mu(q^2) \rangle &=& \frac{1}{2\pi^2}
\langle \bar s(q^2 \gamma^\mu - q^\mu \qslash)P_L b \rangle
\left\{ -\frac{1}{9\epsilon} + \frac43 G(m_c) - \frac23 G(m_b)  
\right\}
- \frac{1}{2\pi^2}
m_b \langle \bar s (\qslash \gamma_\mu - q_\mu) P_R b \rangle \,.
\end{eqnarray}
We denoted here with $G(m_q)$ the function 
appearing in the basic fermion
loop with mass $m_q$\footnote{This function is related to
$h(z,\hat s)$ used in \cite{BuMu} as $h(m_q/m_b,q^2/m_b^2) = -8/3 G(m_q) - 4/9$.} 
\bea
G(m_q) = \int_0^1 \mbox{d} x x(1-x) \log
\left(
\frac{-q^2 x(1-x) + m_q^2 - i\epsilon}{\mu^2}\right)\,.
\eea
In the kinematical region considered here ($4m_c^2 < q^2 < 4m_b^2$), this function is
given explicitly by
\bea
G(m_c) &=& \frac16 \log\left(\frac{m_c^2}{\mu^2}\right) - \frac{5}{18} -
\frac{2m_c^2}{3q^2}
 + \frac16 \sqrt{r}\left(1 + \frac{2m_c^2}{q^2}\right)
\left(\log\frac{1+\sqrt{r}}{1-\sqrt{r}} - i\pi\right)\\
G(0) &=& \frac16 \left[\log\left(\frac{q^2}{\mu^2}\right) -i\pi\right] 
- \frac{5}{18} \\
G(m_b) &=& \frac16 \log\left(\frac{m_b^2}{\mu^2}\right) - \frac{5}{18} -
\frac{2m_b^2}{3q^2}
 + \frac13 \sqrt{\frac{4m_b^2}{q^2}-1}\left(1 + \frac{2m_b^2}{q^2}\right)
\arctan \frac{1}{\sqrt{\frac{4m_b^2}{q^2}-1}}
\eea
where in $G(m_c)$ we denoted $r=\sqrt{1-4m_c^2/q^2}$.

To match onto the operators introduced in Sec.~\ref{sec2} we expand the
results (\ref{loop1})-(\ref{loop6}) in $1/Q$ and go over to the
HQET for the heavy quark field. To the order we work, this amounts to expanding
the charm quark loop using
\bea
G(m_c) = G(0) - \frac{m_c^2}{q^2} + \left(\frac{m_c^2}{q^2}\right)^2
\left[ \log\left(\frac{q^2}{m_c^2}\right) - i\pi - \frac12\right] + \cdots
\eea
On the other hand, since we treat $m_b^2$ and $q^2$ as being comparable,
the full result for the $b$ quark loop function $G(m_b)$ has to be kept.

To illustrate the matching computation we show how the result
(\ref{loop1}) for the T-product containing $Q_1$ is reproduced in 
the operator product expansion (\ref{OPE}). Expanding (\ref{loop1}) in 
powers of $m_c^2/q^2$ one finds 
\bea\label{loopexp}
\langle {\cal T}_1^\mu \rangle &=&
\frac{1}{2\pi^2} [\bar s(q^2 \gamma_\mu -q^\mu \qslash) P_L b]
\left\{
\left[ 4G(0) + \frac23\right]
- 4\frac{m_c^2}{q^2}\right.\\
& &\left.
 + 4 \frac{m_c^4}{q^4}\left[ \log\left(\frac{q^2}{m_c^2}\right)
- i\pi - \frac12\right] + O\left(\frac{m_c^6}{q^6}\right)
\right\} \nonumber
\eea

The terms of $O(q^2)$ and $O(m_c^2)$ in this result can be identified with
the matrix elements of the operators ${\cal O}_1^{(-2)}$ and ${\cal O}_1^{(0)}$,
respectively, provided that their Wilson coefficients are taken to be
\bea
C_{1,1}^{(-2)}(\mu) &=& \frac{1}{2\pi^2}
\left[ 4G(0) + \frac23\right]\,,\qquad
C_{1,1}^{(0)}(\mu) = -\frac{2}{\pi^2}
\eea
Reproducing the $O(m_c^4/q^4)$ term in (\ref{loopexp}) requires the introduction of
dimension-6 operators
containing explicit factors of the charm quark field. They are obtained
by matching from diagrams where the photon attaches to one of the external quark legs
(see Fig.~3).
Expanding these graphs in $1/Q$ and keeping only the term of $O(m_c/Q^2)$
gives (the leading term scales like $\sim 1/Q$, but its $b\to s$ matrix element
vanishes) 
\bea\label{4quark}
{\cal T}_1^\mu(q^2) & &\to {\cal O}^{(2)\mu} =
\frac{8Q_c}{q^2}
\left[ \bar c(\gamma^\nu i\Dright^\mu - i\Dleft^\mu \gamma^\nu) P_L c\right]
(\bar s\gamma_\nu P_L b)\\
& & + \frac{8Q_c}{q^4}  
\left[ \bar c(-\gamma^\nu\qslash\gamma^\mu (q\cdot i\Dright) +
\gamma^\mu \qslash \gamma^\nu (q\cdot i\Dleft)) P_L c\right]
(\bar s\gamma_\nu P_L b)\,.\nonumber
\end{eqnarray}
We dropped here operators which vanish by the equation of motion of the
charm quark field $(i\Dslash - m_c) c = 0$.
\begin{figure}[t!]
\begin{center}
  \includegraphics[width=1.2in]{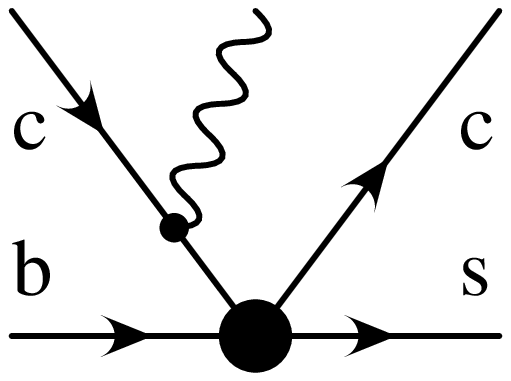}\hspace{0.5cm} +\hspace{0.5cm}
\includegraphics[width=1.2in]{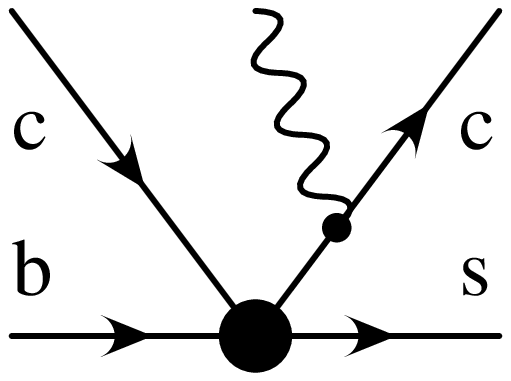} \hspace{0.5cm}$\longrightarrow$
\hspace{0.5cm}
\includegraphics[width=1.2in]{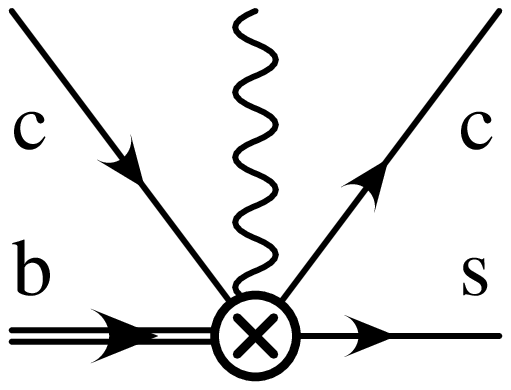}\\
\vspace{0.4cm}
 (a) \hspace{4cm} (b) \hspace{4cm} (c)
\end{center}
{\caption{
Graphs contributing to the matching onto operators with explicit charm fields
(see, e.g. Eq.~(\ref{4quark})). In (a), (b) the filled circle denotes one of the
QCD operators $Q_{1-6}$. The crossed circle in (c) denotes the local operator
appearing in the OPE with quark content $(\bar sb)(\bar cc)$.
The wavy line is the virtual photon $\gamma^*$ connecting to the $e^+e^-$ lepton
pair.}}
\end{figure}
The matrix element of this operator is computed by closing the charm loop, which 
gives
\begin{eqnarray}
\langle s| {\cal O}^{(2)\mu}|b\rangle = 
\frac{N_c Q_c}{\pi^2}\langle \bar s(q^2 \gamma^\mu - q^\mu \qslash )P_L b \rangle
\frac{m_c^4}{q^4}
\left\{ \frac{1}{\epsilon} + \frac32 -
\log\left(\frac{m_c^2}{\mu^2}\right) \right\}
\end{eqnarray}
The coefficient of the logarithmic term $\log m_c$ agrees with that in the
expansion of the exact result in Eq.~(\ref{loopexp}). This shows that the
four-quark operators Eq.~(\ref{4quark}) reproduce the IR of the full theory
result. 
However, these contributions are suppressed by $m_c^4/Q^4 \sim 0.8\%$
relative to those of the leading operators ${\cal O}^{(-2)}_i$, so they can be
expected to be numerically small. This is fortunate, since their matrix elements on 
hadronic states would introduce new unknown form factors in addition
to those contributing to the short-distance amplitude.
In the following  we will not include 4-quark operators 
similar to those in Eq.~(\ref{4quark}).

Using a similar expansion one finds the matching for all remaining 
T-products in (\ref{loop1})-(\ref{loop6}) onto the operators in the OPE
(\ref{OPE}). The results for the Wilson coefficients $C_{i,1}^{(-2,0)}$ are
\begin{eqnarray}\label{Cii}
\frac{1}{C_0^{(v)}(\mu)}
C_{1,1}^{(-2)}(\mu) &=& \frac{1}{2\pi^2}
\left[ 4G(0) + \frac23 \right] - \frac{\alpha_s}{(4\pi)^3} 36C(q^2)\,,\\
\frac{1}{C_0^{(v)}(\mu)}
C_{2,1}^{(-2)}(\mu) &=& \frac{1}{2\pi^2}
\left[\frac43 G(0) + \frac29 \right] - \frac{\alpha_s}{(4\pi)^3} 
(-24B(q^2) + 12C(q^2))\,,\\
C_{3,1}^{(-2)}(\mu) &=&  \frac{1}{2\pi^2}
\left[ \frac{10}{3}G(0) + \frac{1}{27} - \frac83 G(m_b)\right]\,,\\
C_{4,1}^{(-2)}(\mu) &=&  \frac{1}{2\pi^2}
\left[ -\frac23 G(0) - \frac79 - \frac83 G(m_b)\right]\,,\\
C_{5,1}^{(-2)}(\mu) &=&  \frac{1}{2\pi^2}
\left[ 4G(0) - 2 G(m_b) - \frac{7}{27}\right]\,,\\
C_{6,1}^{(-2)}(\mu) &=&  \frac{1}{2\pi^2}
\left[ \frac43 G(0) - \frac23 G(m_b) + \frac19 \right]
\end{eqnarray}
and
\begin{eqnarray}
C_{i,1}^{(0)}(\mu) &=& -\frac{2}{\pi^2}\{ 1\,, \frac13\,, 1\,, \frac13 \,, 1\,, \frac13 \}
\qquad (i = 1-6)
\end{eqnarray}
To facilitate the inclusion of the next-to-leading corrections, these
results were computed using the operator basis in Ref.~\cite{CMM}, and transformed
to the basis in Eq.~(\ref{Qi}) using 4-dimensional Fierz identities.
For this reason, the constant terms in these expressions differ from those in
Eqs.~(\ref{loop1}).
With this convention, the Wilson coefficients $C_i(\mu)$ used in the
remainder of this paper differ beyond the LL approximation from those in
Refs.~\cite{BBL,BuMu} and are equal to the ``barred'' coefficients $\bar C_i(\mu)$
defined in Eq.~(79) of Ref.~\cite{BeFeSe}.
We included here also the next-to-leading results for $C_{1,1}^{(-2)}$
and $C_{2,1}^{(-2)}$, which can be extracted from the recent two-loop
computation of Seidel \cite{Seidel:2004jh} (extending previous approximate
results in \cite{b2see}). The functions $A(s),B(s),C(s)$ are 
given in Eqs.~(29)-(31) of \cite{Seidel:2004jh} and can be written as
\begin{eqnarray}
A(q^2) &=& -\frac{104}{243} \log\frac{m_b^2}{\mu^2} + \delta A(q^2)\nonumber\\
B(q^2) &=& \frac{8}{243}\big[ \big(\frac{4m_b^2}{q^2} - 34 - 17\pi i\big)
\log\frac{m_b^2}{\mu^2} + 8 \log^2\frac{m_b^2}{\mu^2} + 17
\log\frac{q^2}{m_b^2} \log\frac{m_b^2}{\mu^2}\big] \nonumber\\
&-& \frac{16}{243}\big( 1 + \frac{2m_b^2}{q^2} \big) 
\sqrt{\frac{4m_b^2}{q^2}-1} \arctan\frac{1}{\sqrt{\frac{4m_b^2}{q^2}-1}}
\log\frac{m_b^2}{\mu^2} + \delta B(q^2)\nonumber\\
C(q^2) &=&  -\frac{16}{81}\log\frac{q^2}{\mu^2} + \frac{428}{243} -
\frac{64}{27}\zeta(3) + \frac{16}{81}\pi i\nonumber
\end{eqnarray}
The terms $\delta A$ and $\delta B$ do not contain explicit $\mu$ dependence
and take the following values at the zero recoil point in $B\to K^* e^+ e^-$ 
(for $\mu = 4.8$ GeV and $m_b(m_b) = 4.32$ GeV) $\delta A(q^2_{\rm max}) =
0.736 + 0.836 i$,   $\delta B(q^2_{\rm max}) =
-1.332 + 3.058 i$.
$C_0^{(v)}(\mu)$ is one of the Wilson coefficients appearing in the matching 
of the vector current $\bar q\gamma_\mu b$ onto HQET currents and is defined
in Eq.~(\ref{match1}). It accounts for the factorizable two-loop
corrections not included in Ref.~\cite{Seidel:2004jh}.

The results for the coefficients $C_{i,2}^{(-2)}(\mu)$ can be computed in a 
similar way with the results
\begin{eqnarray}
\frac{1}{C_0^{(t)}(\mu)}
C_{1,2}^{(-2)}(\mu) &=& 0(\alpha_s^2)\,,\qquad 
\frac{1}{C_0^{(t)}(\mu)}
C_{2,2}^{(-2)}(\mu) = -\frac{\alpha_s}{(4\pi)^3}(48A(q^2))\,,\\
C_{i,2}^{(-2)}(\mu) &=& \frac{1}{\pi^2}\{ -\frac29\,, -\frac23\,,
\frac{1}{18}\,, \frac16 \}\qquad (i=3-6)\nonumber
\end{eqnarray}
The Wilson coefficient $C_0^{(t)}(\mu)$ appears in the matching of the tensor
current $\bar qi\sigma_{\mu\nu} b$ onto HQET operators and is defined in 
Eq.~(\ref{match2}). The $O(\alpha_s(m_b)$ terms in the first two coefficients have
been extracted from Ref.~\cite{Seidel:2004jh}, where they are given in terms of the 
function $A(q^2)$.

The only dimension-4 operators appearing at this order in matching are 
${\cal O}_{1,4,5}^{(-1)}$, and are introduced through the matching
of the $b$ field onto HQET according to $b = (1+\frac{i\Dslash}{2m_b}) h_v$. 
Their Wilson coefficients are
\bea
C_{i,1}^{(-1)}(m_b) = - C_{i,1}^{(-2)}(m_b)\,,\quad
C_{i,4}^{(-1)}(m_b) =
-C_{i,5}^{(-1)}(m_b) = \frac{1}{2} C_{i,1}^{(-2)}(m_b)\qquad
(i=1-6)\,.
\eea
At two-loop order in the matching, all the other dimension-4 operators will
appear, through the dependence of graphs such as those in Fig.~2(b) on external 
quark momenta.

The gluonic penguin $Q_8$ contributes to the long-distance amplitude
at leading order in $1/Q$ through one-loop graphs. The corresponding
one-loop graphs were computed in the second reference of \cite{b2see} in 
an expansion in $q^2/m_b^2$ and in Ref.~\cite{BeFeSe} for arbitrary $q^2$.
Its contributions to the Wilson coefficients of the leading operators are
\begin{eqnarray}
C_{8,1}^{(-2)}(\mu) = \frac{\alpha_s}{16\pi^3} F_8^{(9)}(q^2)\,,\qquad
C_{8,2}^{(-2)}(\mu) = -\frac{\alpha_s}{8\pi^3} F_8^{(7)}(q^2)
\end{eqnarray}
with $F_8^{(9,7)}(q^2)$ given in Eqs.~(82), (83) of Ref.~\cite{BeFeSe}.
The operator $Q_8$ contributes also at tree level
through gluon-photon scattering graphs (with the
photon coupling to the $b$ and $s$ quarks). Expanding these graphs in 
powers of $1/Q$ one finds at leading order
\begin{eqnarray}
{\cal T}^\mu_8 \to  - \frac{m_b Q_b}{(4\pi)^2 v\cdot q}\bar s_L\sigma_{\alpha\beta}
gG^{\alpha\beta} \gamma^\mu h_{vR}
+ \frac{Q_s}{8 \pi^2}
\bar s_L \gamma_\mu \vslash \sigma_{\alpha\beta}gG^{\alpha\beta} h_{vR}\,.
\end{eqnarray}
The matrix elements of these dimension-6 operators  are
suppressed by $\Lambda^2/Q^2$.

The one-loop graphs in Fig.~2(c) with one insertion of $Q_{1-6}$ produce 
dimension-5 operators containing the gluon
field tensor of the form $\bar s gG_{\mu\nu} h_{v}$. Although their Wilson
coefficients start at $O(\alpha_s^0)$, their matrix elements are $\sim \Lambda^2$,
and therefore are suppressed by $\Lambda^2/Q^2$ relative to the short distance
amplitude. We will neglect all these higher
dimensional operators and keep only the $O(1), O(m_c^2/m_b^2)$ and 
$O(\Lambda/m_b)$ terms in the long distance amplitude.

\section{Matrix elements}

In this section we use the OPE result Eq.~(\ref{OPE}) for the long-distance amplitudes
${\cal T}_i(q^2)$ 
to compute the hadronic amplitude  $A_\mu^{(V)}$ in Eq.~(\ref{AV}) up to
and including corrections of order $O(\alpha_s(Q), \Lambda/m_b, m_c^2/m_b^2)$.
At this point we encounter a technical complication connected with the fact that 
the OPE was performed in terms of HQET operators, while the matrix elements of 
the QCD currents
$\bar s\Gamma b$ appearing in the factorizable matrix elements of $Q_{7,9}$ are
expressed in terms of physical form factors.
This means that the matrix elements of the operators ${\cal O}^{(-2)}_{1,2}$ are
given in terms of HQET form factors, which are not known. Also, keeping all 
$O(\Lambda/m_b)$ contributions requires
that we include also $T$-products of the ${\cal O}^{(-2)}_{1,2}$ operators with
$1/m_b$ sub-leading terms in the HQET Lagrangian. Such nonlocal matrix elements 
introduce additional unknown form factors. This proliferation of unknown matrix elements
appears to preclude a simple form for our final result.

We will show next that it is possible to absorb all these nonlocal 
matrix elements into the physical form factors, through a simple 
reorganization of the operator expansion, such that one is left only with
local $1/m_b$ corrections.
This can be achieved by expressing the leading operators
${\cal O}^{(-2)}_i$ in terms of QCD operators, up to  dimension-4
HQET operators $\bar s iD_\mu (\gamma_5) h_v$. 
Technically, this is obtained by inverting the HQET matching 
 relations (we assume here everywhere the NDR scheme)
\bea\label{match1}
\bar s_L \gamma_\mu b_L &=& C_0^{(v)}(\mu) \bar s_L \gamma_\mu h_{vL} +
C_1^{(v)}(\mu) \bar s_L v_\mu h_{vR} + 
\frac{1}{2m_b} \bar s_L \gamma_\mu i\Dslash h_{vR} + O(1/m_b^2)\\
\label{match2}
\bar s_L i\sigma_{\mu\nu} q^\nu b_R &=& 
C_0^{(t)}(\mu) \bar s_L i\sigma_{\mu\nu} q^\nu h_{vR} +
C_1^{(t)}(\mu) \bar s_L [(v\cdot q) \gamma_\mu - \qslash v_\mu] h_{vL}\\
& & + 
\frac{1}{2m_b} \bar s_L i\sigma_{\mu\nu} q^\nu i\Dslash h_{vL} + O(1/m_b^2)
\,.\nonumber
\eea
The Wilson coefficients $C_i^{(v,t)}(\mu)$ are given 
at one-loop by \cite{EiHi}
\bea
C_0^{(v)}(\mu) &=& 1 - \frac{\alpha_s C_F}{4\pi}(3\log\frac{\mu}{m_b}
+ 4)\,,\qquad 
C_1^{(v)}(\mu) = \frac{\alpha_s C_F}{2\pi}\\
C_0^{(t)}(\mu) &=& 1 - \frac{\alpha_s C_F}{4\pi}(5\log\frac{\mu}{m_b}
+ 4)\,,\qquad 
C_1^{(t)} = O(\alpha^2_s)\,.
\eea
In the $O(1/m_b)$ terms we work at tree level in the matching, which will 
be sufficient for the precision required here, although the method can be
extended to any order in $\alpha_s(m_b)$.

Solving the matching relations Eqs.~(\ref{match1}), (\ref{match2}) for the 
leading order HQET operators appearing in the OPE ${\cal O}_{1,2}^{(-2)}$ one
finds
\begin{eqnarray}\label{inv1}
\bar s_L (q^2 \gamma_\mu - \qslash q_\mu) h_{vL} &=&
\frac{1}{C_0^{(v)}(\mu)} \bar s_L (q^2 \gamma_\mu - \qslash q_\mu) b_L
+ {\cal O}_1^{(-1)} - \frac12 {\cal O}_4^{(-1)} + 
\frac12 {\cal O}_5^{(-1)}\\
\label{inv2}
\bar s_L i\sigma^{\mu\nu} q_\nu h_{vR} &=& \frac{1}{C_0^{(t)}(\mu)}
\bar s_L i\sigma^{\mu\nu} q_\nu b_R - 
\frac{C_1^{(t)}(\mu)}{C_0^{(v)}(\mu)C_0^{(t)}(\mu)}
\bar s_L [(v\cdot q) \gamma_\mu - v_\mu \qslash ] b_L
\end{eqnarray}
We neglected here terms of $O(\alpha_s(m_b) \Lambda/m_b)$.

Substituting these results into the OPE, the leading terms can be written
in terms of physical $B\to K^*$ form factors, with corrections of $O(\Lambda/m_b)$
coming from local dimension-4 operators ${\cal O}_{1-5}^{(-1)}$
\begin{eqnarray}
A_\mu^{(V)} &=& - C_7^{\rm eff}(\mu) \frac{2m_b}{q^2}
\langle \bar s i\sigma_{\mu\nu} q^\nu (1+\gamma_5) b\rangle \\
& & + C_9^{\rm eff}(\mu) 
\langle \bar s \gamma_{\mu}  (1-\gamma_5) b\rangle
+ \frac{1}{q^2} \sum_{i=1}^5 B_i(\mu) \langle {\cal O}_i^{(-1)} \rangle\,.
\nonumber
\end{eqnarray}

We absorbed here the contributions from the leading terms in Eqs.~(\ref{inv1}),
(\ref{inv2}) into 
a redefinition of the Wilson coefficients $C_{7,9}$
\bea\label{C7eff}
C_7(\mu) &\to & C_7^{\rm eff}(\mu) = C_7(\mu) + 2\pi^2\sum_{i=1}^{6,8} C_i(\mu) 
\left[  \frac{C_{i,2}^{(-2)}(\mu)}{C_0^{(t)}(\mu)} +
C_{i,2}^{(0)}(\mu) \frac{m_c^2}{q^2}    \right]\\
\label{C9eff}
C_9(\mu) &\to & C_9^{\rm eff}(\mu) = C_9(\mu)\\
& & - 4\pi^2\sum_{i=1}^{6,8} C_i(\mu) \left[
\frac{C_{i,1}^{(-2)}(\mu)}{C_0^{(v)}(\mu)} -
\frac{C_{i,2}^{(-2)}(\mu)}{C_0^{(t)}(\mu) C_0^{(v)}(\mu)}
C_1^{(t)}(\mu) +
C_{i,1}^{(0)}(\mu) \frac{m_c^2}{q^2}        \right]\,.\nonumber
\eea
The $O(1)$ and $O(m_c^2/q^2)$ contributions to the long-distance amplitude
are contained in $C_{7,9}^{\rm eff}$, and the $O(\Lambda/m_b)$ part is
encoded in the matrix elements of ${\cal O}_i^{(-1)}$. 
Note that the effective Wilson coefficients $C_{7,9}^{\rm eff}$ introduced here
are different from the ``effective Wilson coefficients'' commonly used in the
literature $\tilde C_{7,9}^{\rm eff}$ \cite{BuMu,b2see}. The latter include contributions
from the matrix elements of the
operators $Q_{1-9}$ (usually computed in perturbation theory), and are thus
dependent on the final state. In contrast, our effective Wilson coefficients
are state independent, and encode only contributions from the hard scale
$\mu \sim m_b$.

Combining everything, the next-to-leading expressions for the
effective Wilson coefficients are 
\begin{eqnarray}
C_9^{\rm eff} &=& C_9 -  (C_1 + \frac{C_2}{3})[8G(0)+\frac43] -
C_3 [\frac{20}{3}G(0) - \frac{16}{3} G(m_b) + \frac{2}{27}] +
C_4 [\frac43 G(0) + \frac{16}{3} G(m_b) + \frac{14}{9}]\\ 
&-& 
C_5 [8 G(0) - 4 G(m_b) - \frac{14}{27}] -
C_6 [\frac83 G(0) - \frac43 G(m_b) + \frac{2}{9}] +
\frac{\alpha_s}{4\pi} [C_1 9 C(q^2) + C_2 (-6B(q^2)+3C(q^2))
- C_8 F_8^{(9)}(q^2)]\nonumber\\
C_7^{\rm eff} &=& C_7 -  \frac49 C_3 - \frac43 C_4 + \frac19 C_5 + \frac13 C_6 
+ \frac{\alpha_s}{4\pi}[-C_2 6 A(q^2) - C_8 F_8^{(7)}(q^2)]
\end{eqnarray}

The effective Wilson coefficient $C_9^{\rm eff}$ is RG invariant. At the order
we work here, it satisfies the RG equation
\begin{eqnarray}
\mu\frac{d}{d\mu} C_9^{\rm eff}(\mu) = O(\alpha_s^2 C_{1,2} , \alpha_s C_{3-6})\,.
\end{eqnarray}
The coefficient $C_7^{\rm eff}(\mu)$ satisfies a RG equation 
\begin{eqnarray}\label{gam7}
\mu \frac{\rm d}{{\rm d}\mu} C_7^{\rm eff}(\mu) = \gamma_7(\alpha_s)
 C_7^{\rm eff}(\mu)
\end{eqnarray}
 with anomalous dimension $\gamma_7(\alpha_s) = \gamma_t(\alpha_s)
- \gamma_m(\alpha_s)$ (see Eqs.~(\ref{gamt}) and (\ref{gamm}) for definitions).

The Wilson coefficients of the dimension-4 operators $B_i(\mu)$ are given by
\begin{eqnarray}
B_1(\mu) &=& 8\pi^2 \sum_i C_i(\mu) (C_{i,1}^{(-1)}(\mu) + C_{i,1}^{(-2)}(\mu))\\
B_2(\mu) &=& 8\pi^2 \sum_i C_i(\mu) C_{i,2}^{(-1)}(\mu)\\
B_3(\mu) &=& 8\pi^2 \sum_i C_i(\mu) C_{i,3}^{(-1)}(\mu)\\
B_4(\mu) &=& 8\pi^2 \sum_i C_i(\mu) (C_{i,4}^{(-1)}(\mu) 
-\frac12 C_{i,1}^{(-2)}(\mu))\\
B_5(\mu) &=& 8\pi^2 \sum_i C_i(\mu) (C_{i,5}^{(-1)}(\mu) 
+ \frac12 C_{i,1}^{(-2)}(\mu))\,.
\end{eqnarray}
These Wilson coefficients start at $O(\alpha_s)$ in matching. By absorbing
the factor of $8\pi^2$ in their definition, their expansion in 
$\alpha_s(Q)$ starts with a term of order $\alpha_s(Q)/\pi$.
At the order we work (keeping terms in the OPE of $O(\alpha_s, \Lambda/Q, m_c^2/Q^2)$,
but neglecting $O(\alpha_s \Lambda/Q)$ terms), they all vanish $B_{1-5}=0$.
However, we will include them in the following expressions, which is required
for a complete result to $O(\Lambda/m_b)$ accuracy for the long-distance 
amplitude.

It is convenient to parameterize the physical amplitudes $A_\mu^{(V,A)}$ 
introduced in Eq.~(\ref{AVAdef}) in
terms of eight scalar form factors ${\cal A}^{(V,A)}-{\cal D}^{(V,A)}$ defined as
\bea\label{param}
A_\mu^{(V,A)} &=& {\cal A}^{(V,A)}(q^2) 
i\varepsilon_{\mu\nu\lambda\sigma} \eta^{*\nu}
(p+k)^\lambda (p-k)^\sigma + 
{\cal B}^{(V,A)}(q^2)  \eta_\mu^*\\ 
& & + {\cal C}^{(V,A)}(q^2) 
(\eta^*\cdot p)(p+k)_\mu +
{\cal D}^{(V,A)}(q^2) 
(\eta^*\cdot p)(p-k)_\mu\nonumber
\eea

The $B\to K^* e^+ e^-$ decay rate can be represented as a sum over the
helicity $\lambda=\pm 1, 0$ of the vector meson. In the limit of massless leptons, 
this is given by
\bea
\frac{d\Gamma(B\to K^* e^+ e^-)}{dq^2} = 
\frac{4G_F^2 |V_{tb} V_{ts}^*|^2 \alpha^2}{3m_B^2 (4\pi)^5}
q^2 |\vec q\,| \sum_{\lambda = \pm 1,0}
\left\{ |H_\lambda^{(V)}|^2 + |H_\lambda^{(A)}|^2 \right\}
\eea
where the $H_\lambda^{(V)}$ and $H_\lambda^{(A)}$ correspond to the vector and
axial leptons coupling, respectively.
Expressed in terms of the scalar
amplitudes ${\cal A,B,C}$ introduced in Eq.~(\ref{param}), they are given 
by ($i=V,A$)
\bea
H_\pm^{(i)}(q^2) &=& \mp 2m_B |\vec q| {\cal A}^{(i)}(q^2) - 
{\cal B}^{(i)}(q^2)\\
H_0^{(i)}(q^2) &=& \frac{1}{2m_V\sqrt{q^2}}
\left\{ (-q^2+m_B^2-m_V^2) {\cal B}^{(i)}(q^2) +
4m_B^2 \vec q\,^2 {\cal C}^{(i)}(q^2) \right\}\,.
\eea

The explicit results for the amplitudes ${\cal A}^{(V,A)}-{\cal D}^{(V,A)}$ 
are obtained by taking matrix elements on physical states and are given by
\bea\label{calA}
{\cal A}^{(V)}(q^2) &=& -C_7^{\rm eff}(\mu) \frac{2m_b}{q^2} g_+(q^2) + 
C_9^{\rm eff}(\mu) g(q^2) +  {\cal A}_{\rm l.d.}(q^2)\\
\label{calB}
{\cal B}^{(V)}(q^2) &=& -C_7^{\rm eff}(\mu) \frac{2m_b}{q^2} 
[(m_B^2-m_V^2) g_+(q^2) + q^2 g_-(q^2)]
 - C_9^{\rm eff}(\mu) f(q^2)
+  {\cal B}_{\rm l.d.}(q^2)\\
\label{calC}
{\cal C}^{(V)}(q^2) &=& -C_7^{\rm eff}(\mu) \frac{2m_b}{q^2} 
[-g_+(q^2) + q^2 h(q^2)]
- C_9^{\rm eff}(\mu) a_+(q^2) + 
{\cal C}_{\rm l.d.}(q^2)\\
\label{calD}
{\cal D}^{(V)}(q^2) &=& C_7^{\rm eff}(\mu) \frac{2m_b}{q^2} 
[g_-(q^2) + (m_B^2-m_V^2) h(q^2)]
- C_9^{\rm eff}(\mu) a_-(q^2) + {\cal D}_{\rm l.d.}(q^2)
\eea
and 
\bea
{\cal A}^{(A)}(q^2) &=& C_{10} g(q^2)\\
{\cal B}^{(A)}(q^2) &=& - C_{10} f(q^2)\\
{\cal C}^{(A)}(q^2) &=& - C_{10} a_+(q^2)\\
{\cal D}^{(A)}(q^2) &=& - C_{10} a_-(q^2)\,.
\eea
The coefficients ${\cal D}^{(V,A)}(q^2)$ do not contribute to the 
$B\to V e^+ e^-$ decay rate
into massless leptons, but are relevant for the $B\to V \tau^+ \tau^-$ mode. 
We will not consider them further.
The $O(\Lambda/Q)$ contribution to the long-distance contribution
appears as matrix elements of the
local dimension-4 operators ${\cal O}_i^{(-1)}$ (denoted as
${\cal A}_{\rm l.d.}(q^2)-{\cal D}_{\rm l.d.}(q^2)$ in 
Eqs.~(\ref{calA})-(\ref{calD})). They are given explicitly by
\begin{eqnarray}
{\cal A}_{\rm l.d.}(q^2) &=& \frac{1}{2m_b} d^{(0)}(q^2) B_1 + 
\frac{1}{2m_b} (\bar\Lambda-v.k) g(q^2) B_2
- \frac{1}{4m_b} [(1+\frac{\bar\Lambda}{m_B}) g_+(q^2) +
(1-\frac{\bar\Lambda}{m_B}) g_-(q^2)] B_4\\
\nonumber & & + \frac{m_s}{2m_b} g(q^2) B_5\\
{\cal B}_{\rm l.d.}(q^2) &=& \frac{1}{2m_b} d_1^{(0)}(q^2) B_1 - \frac{1}{2m_b} 
(\bar\Lambda-v.k) f(q^2) B_2\\
\nonumber
& & - \frac{1}{2m_b}
[(\bar\Lambda v-k).(p+k) g_+(q^2) + (\bar\Lambda v-k).(p-k) g_-(q^2)] B_4
+ \frac{m_s}{2m_b} f(q^2) B_5\\
{\cal C}_{\rm l.d.}(q^2) &=&
\frac{1}{2m_b} [d_+^{(0)}(q^2)-\frac{\bar\Lambda-v.k}{2m_b} s(q^2)] B_1 -
\frac{1}{2m_b}(\bar\Lambda - v.k)
 [ a_+(q^2) + \frac{1}{2m_b} s(q^2)] B_2\\
\nonumber
& & - \frac{1}{4m_b} (1-\frac{v.k}{m_B}) s(q^2) B_3 + 
(\frac{\bar\Lambda}{2m_b^2} g_+(q^2) - \frac12 (\bar\Lambda - v.k) h(q^2)
- \frac{1}{4m_b} s(q^2)) B_4  \\
\nonumber & & +
\frac{m_s}{2m_b} 
[ a_+(q^2) + \frac{1}{2m_b} s(q^2)] B_5
\end{eqnarray}
The form factors appearing here are defined in the Appendix. The corresponding
result for ${\cal D}_{\rm l.d.}(q^2)$ can be obtained from the Ward identity
which gives ${\cal B}_{\rm l.d.}(q^2) + (m_B^2-m_V^2) {\cal C}_{\rm l.d.}(q^2)
 + q^2 {\cal D}_{\rm l.d.}(q^2) = 0$.
Expanding in powers of $1/m_b$ and keeping the leading terms gives
${\cal D}_{\rm l.d.}(q^2) = - {\cal C}_{\rm l.d.}(q^2) + O(\Lambda/m_b)$.

For completeness we quote here also the relevant results for the
semileptonic decay $B\to \rho e\bar\nu$.
The decay rate is given by a sum over contributions corresponding to 
helicities of the final vector meson $\lambda = \pm,0$
\bea
\frac{d\Gamma(\bar B\to \rho e \nu)}{dq^2} = 
\frac{G_F^2 |V_{ub}|^2}{96 \pi^3 m_B^2}
q^2 |\vec q\,| \sum_{\lambda = \pm 1,0} |H_\lambda|^2
\eea
where the helicity amplitudes are given by
\begin{eqnarray}
H_\pm(q^2) &=& \mp 2m_B |\vec q| g(q^2) + f(q^2)\\ 
H_0(q^2) &=& \frac{1}{2m_V\sqrt{q^2}}
\left\{ (q^2-m_B^2+m_V^2) f(q^2) -
4m_B^2 \vec q\,^2 a_+(q^2) \right\}\,.
\end{eqnarray}

\section{Phenomenology}

In the low recoil region, the amplitudes ${\cal A}^{(V)},{\cal B}^{(V)},
{\cal C}^{(V)}$  for $B\to K^*\ell^+\ell^-$ are dominated by the operator 
$Q_9$. The contribution proportional to $C_7$ can be
expressed in terms of the $C_9$ terms using the form factor relations
Eqs.~(\ref{W-1p}),(\ref{W-2p}),(\ref{W-3p}) given in the Appendix. 
Keeping terms to subleading order in $\Lambda/m_b$, these amplitudes
can be written as
\begin{eqnarray}
{\cal A}^{(V)}(q^2) &=& C_9^{\rm eff} g(q^2) \left\{
1 + \frac{C_7^{\rm eff}}{C_9^{\rm eff}} \frac{2m_b}{q^2}
\left[ (1+\frac{2D_0^{(v)}(\mu)}{C_0^{(v)}(\mu)}) m_b + m_q + 2 
\frac{d^{(0)}(q^2)}{g(q^2)} \right] + 
\frac{{\cal A}_{\rm l.d.}(q^2)}{C_9^{\rm eff} g(q^2)} +
O(\frac{\Lambda^2}{m_b^2})
\right\}\\
{\cal B}^{(V)}(q^2) &=& -C_9^{\rm eff} f(q^2) \left\{
1 + \frac{C_7^{\rm eff}}{C_9^{\rm eff}} \frac{2m_b}{q^2}
\left[ (1+\frac{2D_0^{(v)}(\mu)}{C_0^{(v)}(\mu)}) m_b - m_q - 
2 \frac{d_1^{(0)}(q^2)}{f(q^2)} \right] -
\frac{{\cal B}_{\rm l.d.}(q^2)}{C_9^{\rm eff} f(q^2)} +
O(\frac{\Lambda^2}{m_b^2})
\right\}\\
{\cal C}^{(V)}(q^2) &=& -C_9^{\rm eff} a_+(q^2) \left\{
1 + \frac{C_7^{\rm eff}}{C_9^{\rm eff}} \frac{2m_b}{q^2}
\left[(1+\frac{2D_0^{(v)}(\mu)}{C_0^{(v)}(\mu)}) m_b - m_q - 
2 \frac{d_+^{(0)}(q^2)}{a_+(q^2)} \right] + 
\frac{{\cal C}_{\rm l.d.}(q^2)}{C_9^{\rm eff} a_+(q^2)} +
O(\frac{\Lambda^2}{m_b^2}) \right\}\,.
\end{eqnarray}

Inserting these results into the expressions for the helicity 
amplitudes $H_\lambda^{(V)}(q^2)$ one finds
\begin{eqnarray}
H_\pm^{(V)}(q^2) &=& \mp 2m_B m_V \sqrt{y^2-1} C_9^{\rm eff} g(q^2) 
(1 + \delta + r_a)
+ C_9^{\rm eff} f(q^2) (1 + \delta + r_b)\\
H_0^{(V)}(q^2) &=& - \frac{m_B y - m_V}{\sqrt{q^2}} C_9^{\rm eff} f(q^2) 
(1+\delta + r_b)
- 2m_B^2 m_V \frac{y^2-1}{\sqrt{q^2}} C_9^{\rm eff} a_+(q^2) (1 + \delta + r_c)
\end{eqnarray}
Here $1+\delta(q^2)$ scales like $m_b^0$ and $r_{a,b,c}(q^2)$ parameterize 
the $1/m_b$ correction. Their explicit expressions are
\begin{eqnarray}
\delta(q^2) &=& \frac{C_7^{\rm eff}(\mu)}{C_9^{\rm eff}} \frac{2m_b^2(\mu)}{q^2}
\left(1 + \frac{2 D_0^{(v)}(\mu)}{C_0^{(v)}(\mu)}\right)\\
r_a(q^2) &=& \delta(q^2) \frac{1}{m_b}(m_q + 2 \frac{d^{(0)}(q^2)}{g(q^2)}) 
+\frac{{\cal A}_{\rm l.d.}(q^2)}{C_9^{\rm eff} g(q^2)}  \\
\label{rb}
r_b(q^2) &=& \delta(q^2) \frac{1}{m_b} ( - m_q - 2 
\frac{d_1^{(0)}(q^2)}{f(q^2)}) +
\frac{{\cal B}_{\rm l.d.}(q^2)}{C_9^{\rm eff} f(q^2)}  \\
r_c(q^2) &=& \delta(q^2) \frac{1}{m_b} (- m_q - 2 
\frac{d_+^{(0)}(q^2)}{a_+(q^2)}) -
\frac{{\cal C}_{\rm l.d.}(q^2)}{C_9^{\rm eff} a_+(q^2)}\,.
\end{eqnarray}
Combining the RG equations satisfied by $C_7^{\rm eff}(\mu)$ Eq.~(\ref{gam7})
and by the $1 + 2 D_0^{(v)}(\mu)/C_0^{(v)}(\mu)$ factor Eq.~(\ref{gamD}), one can see that the
$\delta(q^2)$ parameter is RG invariant.

These results imply that {\em the $H_\lambda^{(V)}(q^2)$ amplitudes
for rare $B\to V \ell^+\ell^-$ decays are related at leading order 
in $\Lambda/m_b$  to those
for semileptonic decay $B\to V e\bar\nu$ with a common proportionality
factor}
\begin{eqnarray}\label{helrel}
H_\lambda^{(V)}(q^2) = C_9^{\rm eff} 
(1 + \delta(q^2) + O(\Lambda/m_b)) H_\lambda(q^2)\,.
\end{eqnarray}
Combining this with the rate formulas one finds a
relation among the decay rates for the rare and semileptonic decays
\begin{eqnarray}\label{11}
\frac{\mbox{d}\Gamma(\bar B\to \rho e\nu)/\mbox{d}q^2}
{\mbox{d}\Gamma(\bar B\to K^* \ell^+ \ell^-)/\mbox{d}q^2}
= 
\frac{|V_{ub}|^2}{|V_{tb} V_{ts}^*|^2} \cdot \frac{8\pi^2}{\alpha^2}\cdot
\frac{1}{|C_9^{\rm eff}(1+\delta(q^2))|^2 + |C_{10}|^2}
\frac{\sum_\lambda |H_\lambda^{B\to\rho}(q^2)|^2}
{\sum_\lambda |H_\lambda^{B\to K^*}(q^2)|^2}
\end{eqnarray}
The corrections to this relation are of order $O(\Lambda/m_b)$ and can be
expressed in terms of the three parameters $r_{a,b,c}(q^2)$ introduced above.

The ratio of decay rates in Eq.~(\ref{11}) has been considered previously
in Ref.~\cite{SaYa,LiWi,LSW} in connection with a method for determining 
$|V_{ub}|$. This requires some information about the SU(3) breaking ratio
of helicity amplitudes appearing on the right-hand side 
\begin{eqnarray}
R_B(y) \equiv \frac{\sum_\lambda |H_\lambda^{B\to\rho}(y)|^2}
{\sum_\lambda |H_\lambda^{B\to K^*}(y)|^2} 
\end{eqnarray}
It has been proposed in \cite{LiWi,LSW} to 
determine $R_B$ in terms of the corresponding ratio of $D\to \rho/K^*$ 
decay amplitudes $R_D(y)$ using a double ratio \cite{Grin}, up to corrections 
linear in both heavy quark and SU(3) symmetry breaking
\begin{eqnarray}\label{dbl}
R_B(y) = R_D(y)(1 + O(m_s(\frac{1}{m_c}-\frac{1}{m_b}))
\end{eqnarray}
In this relation, the two sides must be taken at the same value of the
kinematical variable $y=E_V/m_V$.
A chiral perturbation theory computation \cite{LSW} at the zero recoil point 
$y=1$ shows
that the corrections to this prediction are even smaller than suggested by
the naive dimensional estimate Eq.~(\ref{dbl}). We do not have anything new to 
add on this point, and focus instead on the structure of the denominator in 
Eq.~(\ref{11}).

The results of our paper improve on previous work in two main respects. 
First, we point out that the rate ratio (\ref{11}) can be computed at leading 
order in
$1/m_b$ over the entire small recoil region, and not only at the zero recoil
point $q^2 = (m_B-m_V)^2$. 
This has important experimental implications, as the rate itself vanishes at 
the zero recoil point,
such that measuring the ratio in Eq.~(\ref{11}) would  involve an 
extrapolation from
$q^2 < q^2_{\rm max}$. Most importantly,  Eq.~(\ref{11}) allows the 
determination of $V_{ub}$ using ratios of rates integrated over a range
in $q^2$, as long as such a range is still contained within the low recoil region.

Second, we present
explicit results for the subleading $O(\Lambda/m_b)$ correction to this
result in terms of new form factors contained in the parameters 
$r_{a,b,c}(q^2)$. Using model computations of these form factors, this allows
a quantitative estimate of the power corrections effect on the 
$V_{ub}$ determination.

In the rest of this section we will study in some detail the (RG-invariant) 
quantity $N_{\rm eff}(q^2)$ defined through the 
ratio of rare radiative and semileptonic decays in Eq.~(\ref{11})
\begin{eqnarray}
\frac{\mbox{d}\Gamma(\bar B\to \rho e\nu)/\mbox{d}q^2}
{\mbox{d}\Gamma(\bar B\to K^* \ell^+ \ell^-)/\mbox{d}q^2}
= 
\frac{|V_{ub}|^2}{|V_{tb} V_{ts}^*|^2} \cdot \frac{8\pi^2}{\alpha^2}\cdot
\frac{1}{N_{\rm eff}(q^2)}
\frac{\sum_\lambda |H_\lambda^{B\to\rho}(q^2)|^2}
{\sum_\lambda |H_\lambda^{B\to K^*}(q^2)|^2}
\end{eqnarray}
The results of this paper offer a systematic way of computing this
quantity in an expansion in $\alpha_s(Q)$, $m_c^2/Q^2$ and $\Lambda/m_b$.
The precision of a $|V_{ub}|$ determination using this method is ultimately
determined by the precision in our knowledge of this parameter. 
There are several sources of uncertainty in $N_{\rm eff}(q^2)$, coming from 
scale dependence,  $O(\Lambda/m_b)$ power 
corrections and duality violations. We will consider them  in turn.

At leading order in $\Lambda/m_b$, the $N_{\rm eff}(q^2)$ parameter is given by
\begin{eqnarray}\label{NeffLO}
N_{\rm eff}(q^2) = |C_9^{\rm eff} + \frac{2m_b(\mu)^2}{q^2} C_7^{\rm eff}
\left(1 + 2\frac{D_0^{(v)}(\mu)}{C_0^{(v)}(\mu)}\right)|^2 + 
|C_{10}|^2 +  O(\Lambda/m_b)\,.
\end{eqnarray}
We give in Table II results for the effective Wilson coefficients
$C_{7,9}^{\rm eff}$ at several values of the renormalization scale
$\mu \sim m_b$. We work both at leading log order (next-to-leading log
order for $C_9(\mu)$), and at next-to-leading order (NNLL order for $C_9$).
In each of these approximations the combination of effective Wilson coefficients
in Eq.~(\ref{NeffLO}) satisfies the RG equation
\begin{eqnarray}
\mu\frac{d}{d\mu}\left[ C_9^{\rm eff} + \frac{2m_b(\mu)^2}{q^2} C_7^{\rm eff}
\left(1 + 2\frac{D_0^{(v)}(\mu)}{C_0^{(v)}(\mu)}\right) \right] =
\left\{
\begin{array}{cc}
O(\alpha_s)\,, & \mbox{(LL)} \\
O(\alpha_s^2 C_{1,2}, \alpha_s C_{3-6})\,, & \mbox{(NLL)} \\
\end{array}
\right.
\end{eqnarray}

The structure of the NNLL running for the Wilson coefficients in the 
$b\to s e^+ e^-$ weak Hamiltonian was given in Ref.~\cite{BoMiUr} (see also 
\cite{BeFeSe}). The complete NNLL result requires  the 3-loop 
mixing of the
four-quark operators into $Q_{7,9}$, which was obtained only recently \cite{ADM}.
We use here the full NNLL results for the Wilson coefficients $C_{7,9}$, 
which were presented in \cite{BoGaGoHa}.
The factor containing $D_0^{(v)}(\mu)$ can be extracted from Eq.~(\ref{kappa1}) 
and its 
inclusion is necessary at NLL to achieve the scale independence of $N_{\rm eff}$
to this order.

\begin{table}[t!]
\begin{center}
\begin{tabular}{c|c|cc|cc|cc|cc}
\hline\hline
 & $\mu_b$ (GeV) & $C_9$ & $C_7$ & $C_9^{\rm eff}(y=1)$ & $C_9^{\rm eff}(y=1.5)$ & 
$C_7^{\rm eff}(y=1)$ & $C_7^{\rm eff}(y=1.5)$ & 
$N_{\rm eff}(y=1)$ & $N_{\rm eff}(y=1.5)$\\
\hline
   & 2.4 & 4.378 & -0.388 & $4.315 + 0.198 i$ & $4.338 + 0.198 i$ 
   & $C_7$ & $C_7$ & 30.80 & 28.96 \\
LL & 4.8 & 4.140 & -0.343 & $4.331 + 0.550 i$ & $4.395 + 0.550 i$ 
   & $C_7$ & $C_7$ & 33.37 & 32.34 \\
   & 9.6 & 3.760 & -0.304 & $4.420 + 0.822 i$ & $4.513 + 0.822 i$ 
   & $C_7$ & $C_7$ & 35.81 & 35.38 \\
\hline\hline
& 2.4 & 4.510 & -0.366 & $4.685 + 0.494 i$ & $4.742 +0.442 i$ 
  & -0.352-0.127i & -0.360-0.122i & 32.75 & 30.83 \\
NLL & 4.8 & 4.218 & -0.332 & $4.611 + 0.556 i$ & $4.680 + 0.514 i$ 
  & -0.401-0.100i & -0.408-0.097i & 32.76 & 31.11 \\
& 9.6 & 3.799 & -0.300 & $4.589 + 0.643 i$ & $4.668 + 0.609 i$ 
  & -0.422-0.083i & -0.428-0.080i & 33.46 & 32.10 \\
\hline\hline
\end{tabular}
\end{center}
\caption{\setlength\baselineskip{12pt}
Results for  the Wilson coefficients in the weak Hamiltonian $C_{7,9}$
and the effective Wilson coefficients appearing in the $B\to K^* e^+e^-$
decay rate at LL and NLL order. 
The Wilson coefficient
$C_{10}$ is equal to $C_{10}^{\rm NLL} = -4.409$ and 
$C_{10}^{\rm NNLL} = -4.279$. The other parameters used here are
$m_b(m_b)=4.32$ GeV, $\alpha_s(M_Z)=0.119$ and $m_c(m_c)=1.335$ GeV.
}
\end{table}

To illustrate the $q^2$ dependence of the effective Wilson coefficients,
we quote their values at two kinematical points $y=1$ and $y=1.5$, 
corresponding to the low recoil
region overlapping with that kinematically accessible in $D$ decays. The
resulting dependence on $y$ is very mild, of about 2.5\% in $C_9^{\rm eff}$
and almost negligible in $C_7^{\rm eff}$. 

Next we consider the scale dependence of the results, by computing
the variation of the effective Wilson coefficients 
between the scales $2\mu_b$ and $\mu_b/2$ with $\mu_b = 4.8$ GeV.
The LLO Wilson coefficient $C_9$ changes in this range by 15\%, while the corresponding
variation in $C_9^{\rm eff}$ is reduced to 2\% (for the real part), and 36\%
(for $\frac{1}{\pi}\mbox{Im }C_9^{\rm eff}$).
At NNLL the change in $C_9$ is 17\%, which is reduced in the effective
Wilson coefficient $C_9^{\rm eff}$ to 2\% for $\mbox{Im }C_9^{\rm eff}$, 
and $8.5\%$ for $\frac{1}{\pi}\mbox{Im }C_9^{\rm eff}$.
Combining everything, at LL order the scale dependence of $N_{\rm eff}$ is about 16\% 
which is reduced at NNL order to about $3.5\%$ (at the zero recoil point $y=1$). 

To get a sense for the relative contributions
to  the long-distance effects in $C_9^{\rm eff}$, we give below
the detailed structure of this effective coefficient at LL and NLL orders for 
$\mu_b=4.8$ GeV at $y=1$
\begin{eqnarray}
LL &:& C_9^{\rm eff}(y=1) = 4.140 + (0.136 + 0.506 i) + (0.004 + 0.044 i) 
+ 0.000 + 0.050
= 4.330 + 0.550 i\\
NLL &:& C_9^{\rm eff}(y=1) = 4.218 + (0.313 + 0.505 i) + (0.001 + 0.050 i) 
- 0.006 + 0.085
= 4.611 + 0.556 i\,.\nonumber
\end{eqnarray}
The five terms  correspond to  $C_9$, the contribution of
$Q_{1,2}$, from $Q_{3-6}$, $Q_8$ and the $m_c^2/Q^2$ term respectively.
As expected, the dominant contribution to the long-distance part of
$C_9^{\rm eff}$ comes from the operators $Q_{1,2}$, with $Q_{3-6}$ contributing
about 3\% and the $m_c^2/Q^2$ term about 0.1\%.

The structure of the power corrections of $O(\Lambda/m_b)$ is in general very 
complicated and depends on both the leading and subleading $B\to V$ form factors.
Details of such an analysis will be presented
elsewhere. We will limit ourselves here to the study of these corrections
at the zero recoil point, where they are given only by $r_b(q^2)$, defined in
Eq.~(\ref{rb}). At the
zero recoil point $q^2=q_{\rm max}^2$, the relation among rare radiative
and semileptonic helicity amplitudes Eq.~(\ref{helrel})
can be extended to subleading order in $1/m_b$ and reads
\begin{eqnarray}
H_\lambda^{(V)}(q^2_{\rm max}) =
C_9^{\rm eff} (1+\delta(q^2_{\rm max}) + r_b(q^2_{\rm max})) 
H_\lambda(q^2_{\rm max})\,.
\end{eqnarray}
The corresponding modification of the relation for decay rates 
Eq.~(\ref{11}) is obtained
by the replacement $1+\delta(q^2) \to 1+\delta(q^2) +r_b(q^2)$. 
Since the leading order result for $N_{\rm eff}(q^2)$ has only a weak 
dependence on $q^2$ in the low recoil region
(see Table II), this is a reasonably good approximation.

A complete computation of $r_b(q^2_{\rm max})$ is not possible at present
as ${\cal B}_{\rm l.d.}$ depends on the (as yet unknown) Wilson coefficients 
$B_{1-5}$. Dimensional analysis estimates of the first term in (\ref{rb}) give
$r_b(q^2_{\rm max}) \sim -(0.03 \pm 0.01) \Lambda/m_b$, which represents
at most an uncertainty of $1\%$ in $N_{\rm eff}(q^2_{\rm max})$.
Barring an anomalously large value for ${\cal B}_{\rm l.d.}$, this
suggests very small power corrections to the coefficient $N_{\rm eff}$.

Finally, we address the issue of duality violations. Their effects are 
difficult to quantify in a precise way, but some guidance can be obtained 
from the experimental data on the 
$R=\sigma(e^+e^-\to hadrons)/\sigma(e^+e^-\to \mu^+\mu^-)$
ratio, to which the coefficient $N_{\rm eff}(q^2)$ is very similar.
Good data is available on the ratio $R$ in the $c\bar c$ resonance 
region (see e.g. Fig.~39.8 in \cite{PDG}). In the region $\sqrt{q^2} =
4.1-4.4$ GeV (corresponding to the kinematics relevant here), the ratio 
$R$ oscillates around its pQCD predicted value by less than $\sim 25\%$. 
Strictly speaking, the quantity analogous to $R$ in our case is Im$(C_9^{\rm eff})$,
which represents only about 12\% of the magnitude of $|C_9^{\rm eff}|$. In
the real part of $C_9^{\rm eff}$, the relative error introduced by these 
oscillations is
suppressed by the large value of $C_9$ to about $0.3/4.3\times 10\% \sim
1\%$. Due to the fact that Im $(C_9^{\rm eff}$)/Re $(C_9^{\rm eff}) \sim 12\%$, the 
25\% duality violation effect in Im$(C_9^{\rm eff})$ is reduced in
$|C_9^{\rm eff} + 2m_q^2/q^2 C_7^{\rm eff}|^2$ to about 2\%.
The corresponding effect in $N_{\rm eff}$ is reduced by a further factor of 0.5
since the contributions of the two terms in $N_{\rm eff}$ are roughly equal,
and $C_{10}$ is an invariant.
These arguments show that duality violation effects are likely to be very
small in $N_{\rm eff}$ in the kinematical region considered, 
probably below 5\%.
Precise measurements of the $q^2$ spectrum in this region could help resolve 
and reduce this source of uncertainty.

Combining all sources of errors, we find a total uncertainty in 
$N_{\rm eff}$ of less than $\sim 10\%$, which is dominated by duality
violation effects. This gives a total theory uncertainty 
on $|V_{ub}|$ from this method of about 5\%.

We comment briefly on the experimental feasibility of this method.
Model estimates of the dilepton invariant mass spectrum in $B\to K^* \mu^+\mu^-$
indicate that the integrated branching ratio corresponding to the
region considered here $q^2 = [15,19]$ GeV$^2$ is about $(2-5)\times 10^{-7}$,
depending on the form factor models used \cite{ABHH}. Extrapolating the uncertainties
in the present data \cite{Babar,Belle} to 1000 fb$^{-1}$, corresponding to the
entire data sample from the B factories, suggests that this
integrated branching ratio will be measured to about 25\%. This is
beginning to be comparable to the theory uncertainty, and indicates
that a competitive determination of $|V_{ub}|$ using this method
will likely require a super B-factory.

\section{Conclusions}

We presented in this paper a short-distance expansion for the long-distance
contributions to exclusive $B\to K^{(*)} \ell^+ \ell^-$ decays in the small
recoil region. The main observation is that in this kinematical region,
there are 3 relevant energy scales: $Q = m_b \sim \sqrt{q^2}, m_c, \Lambda$.
We use an operator product expansion (OPE) and the heavy quark effective
theory (HQET) to integrate out the effects of the large scale $Q$, and 
classify the effects from the remaining scales in terms of operators 
contributing at a given order in $m_c^2/Q^2$ and $\Lambda/Q$.

Our main result is a systematic expansion for the long-distance amplitude
in $B\to K^{(*)} \ell^+ \ell^-$ decays
including terms of $O(m_c^2/Q^2)$ and $O(\Lambda/Q)$, which can be
extended to any order in $\alpha_s(Q)$. The final results for physical
observables are explicitly scale and scheme independent, order by order
in perturbation theory. This is to be contrasted with the often used
naive factorization approximation (combined with resonance saturation),
which is not consistent with constraints imposed by renormalization group
evolution.

The form of the result is analogous to that for the
$R$ ratio in $e^+e^-\to hadrons$, which can be computed systematically
in an expansion in $1/Q^2$. For example, the nonperturbative effects in
the $R$ ratio have an analog in the $b\to s e^+ e^-$ case as form factors 
of higher dimensional flavor-changing currents. We classify all the
nonperturbative matrix elements required for a complete description of
$B\to K^{(*)} \ell^+ \ell^-$ to the order considered.
We find that none of these new form factors enter
at order $O(1)$ and $O(m_c^2/m_b^2)$ for the long-distance contribution,
and start contributing first at $O(\alpha_s(Q) \Lambda/m_b)$.

These results are applied to a method for extracting the CKM matrix 
element $V_{ub}$ from the ratio of semileptonic and rare exclusive $B$ 
decays in the small recoil region. We find that the long-distance
effect in this determination is well controlled by the expansion in
$\Lambda/m_b$ and $m_c^2/m_b^2$, and the precision of such a method is
dominated by scale dependence and
duality violating effects. Experimental measurements of the dilepton
invariant mass spectrum $d\Gamma/dq^2$ in $B\to K^{(*)} \ell^+ \ell^-$
will allow a direct control of these effects.

The methods of this paper can be applied to other problems of interest for
the phenomenology of rare B decays. The long distance amplitude has a 
complex phase, which is
however completely calculable using the OPE. This means that observables
such as CP violating asymmetries (in the SM and beyond) can be computed 
in a model-independent way. 
Combined with methods based on the soft-collinear effective theory (SCET)
\cite{SCET} and perturbative QCD \cite{BeFe,BeFeSe}, which are applicable at 
large recoil, the approach proposed here
opens up the possibility of attacking the exclusive 
$b\to s e^+ e^-$ rare $B$ decays from the both ends of the $q^2$ spectrum.

\begin{acknowledgments}
We would like to thank Christoph Bobeth for providing us with the Mathematica
code for the NNLO running of the Wilson coefficients presented in 
Ref.~\cite{BoGaGoHa}. D.P. is grateful to Andrzej Czarnecki for useful discussions.
The work of B.G. was supported in part by the Department of Energy under Grant 
DE-FG03-97ER40546.
The work of D. P. has been supported by the U.S. Department of Energy (DOE)
under the Grant No. DF-FC02-94ER40818.
\end{acknowledgments}

\newpage
\appendix
\section{Form factor relations}

We give here an alternative derivation of the improved heavy quark symmetry
form factor relations at low recoil presented in Ref.~\cite{GrPi1}, including
the leading power corrections $\sim O(\Lambda/m_b)$ and hard gluon effects.
As a by-product we derive exact relations for the HQET Wilson coefficients
of dimension-4 operators following from the equations of motion.

We start by giving the definitions of the $B\to V$ form factors used. 
One possible parameterization is
\begin{eqnarray}\label{Vdef}
\langle V(k,\eta)|\bar q\gamma_\mu b|\bar B(p)\rangle &=&
g(q^2) i\varepsilon_{\mu\nu\lambda\sigma} \eta^{*\nu}
(p+k)^\lambda (p-k)^\sigma\\
\label{Adef}
\langle V(k,\eta)|\bar q\gamma_\mu\gamma_5 b| \bar B(p)\rangle &=&
f(q^2) \eta^*_\mu 
+ \, a_+(q^2)(\eta^*\cdot p)(p+k)_\mu\\
 & &\qquad\quad\,\,\,\, +\,
a_-(q^2)(\eta^*\cdot p)(p-k)_\mu\,,\nonumber\\
\label{Tdef}
\langle V(k,\eta)|\bar qi\sigma_{\mu\nu} b|\bar B(p)\rangle &=&
g_+(q^2) i\varepsilon_{\mu\nu\lambda\sigma} \eta^{*\lambda}
(p+k)^\sigma + 
g_-(q^2) i\varepsilon_{\mu\nu\lambda\sigma} 
\eta^{*\lambda} (p-k)^\sigma\\
 & +&
h(q^2)(\eta^*\cdot p) i\varepsilon_{\mu\nu\lambda\sigma} (p+k)^\lambda
(p-k)^\sigma \nonumber\,.
\eea
We use the convention $\varepsilon^{0123}=1$. This particular
definition of the form factors is convenient in the low recoil region
$q^2 \sim (m_B-m_V)^2$, where it
simplifies the power counting in $m_b$. Taking into  account the usual 
relativistic normalization of the $B$ meson state, these form factors 
satisfy the scaling laws \cite{IsWi}
\bea\nonumber
& & f(q^2) \propto m_b^{1/2}\,,\quad
g(q^2) \propto m_b^{-1/2}\,,\quad
a_+(q^2)-a_-(q^2) \propto m_b^{-1/2}\,,\quad
a_+(q^2)+a_-(q^2) \propto m_b^{-3/2}
\\
\label{pc}
& & g_+(q^2) - g_-(q^2) \propto m_b^{1/2}\,,\qquad 
g_+(q^2) + g_-(q^2) \propto m_b^{-1/2}\,,\qquad
h(q^2) \propto m_b^{-3/2}
\,.
\eea 
We will require also the form factor of the pseudoscalar density 
defined as
\begin{eqnarray}
\langle V(k,\eta)|\bar q\gamma_5 b|\bar B(p)\rangle &=&
(\eta^*\cdot p) s(q^2)\,.
\end{eqnarray}
This is not independent and can be obtained using the equation of motion
for the quark fields in terms of the form
factors defined above as
\begin{eqnarray}
s(q^2) = -\frac{1}{m_b+m_q}[f(q^2) + (m_B^2-m_V^2) a_+(q^2) + q^2 a_-(q^2)]\,.
\end{eqnarray}
The leading term in the expansion of $s(q^2)$ in powers of $\Lambda/m_b$
scales like $s(q^2) \propto m_b^{-1/2}$ and can be written as
\begin{eqnarray}
s(q^2) = -\frac{1}{m_B}f(q^2) - a_+(q^2) (m_B+v\cdot k) - 
a_-(q^2) (m_B-v\cdot k) + O(m_b^{-3/2})\,.
\end{eqnarray}

An alternative parameterization commonly used in the literature 
defines the form factors as (with $q_\mu = p_\mu - k_\mu$)
\bea
\langle V(k,\eta)|\bar q\gamma_\mu(1-\gamma_5) b|\bar B(p)\rangle &=& 
\frac{2V(q^2)}{m_B+m_V} \,i\varepsilon_{\mu\nu\rho\sigma}
 \eta^{\ast\nu} \, p^\rho\: k^\sigma \\
& &\hspace*{-2cm} - 2m_VA_0(q^2)\,\frac{\eta^\ast\cdot p}{q^2}\,q_\mu - 
  (m_B+m_V)\,A_1(q^2)\left[\eta^{\ast\mu}-
  \frac{\eta^\ast\cdot p}{q^2}\,q^\mu\right] \nonumber \\
& &\hspace*{-2cm}
+\,A_2(q^2)\,\frac{\eta^\ast\cdot p}{m_B+m_V}
 \left[p_\mu+k_{\mu} -\frac{m_B^2-m_V^2}{q^2}\,q_\mu\right],\nonumber \\
\langle V(k,\eta)|\bar qi\sigma_{\mu\nu} q^\nu b|\bar B(p)\rangle &=& 
-2T_1(q^2) \,i\varepsilon_{\mu\nu\rho\sigma}
 \eta^{\ast\nu} \, p^\rho\: k^\sigma\\
\langle V(k,\eta)|\bar qi\sigma_{\mu\nu} q^\nu \gamma_5 b|\bar B(p)\rangle &=& 
T_2(q^2) [(m_b^2-m_V^2) \eta_\mu^\ast - (\eta^*\cdot p)(p_\mu+k_\mu)]\\
& & +
T_3(q^2) \frac{\eta^*\cdot p}{m_B^2-m_V^2} [(m_B^2-m_V^2) (p_\mu - k_\mu) - q^2
(p_\mu + k_\mu)]\nonumber
\eea
The relation to the alternative definition in Eqs.~(\ref{Vdef})-(\ref{Tdef}) is
\begin{eqnarray}
g(q^2) &=& - \frac{1}{m_B+m_V} V(q^2)\,,\qquad f(q^2) = (m_B + m_V) A_1(q^2) \\
a_+(q^2) &=& -\frac{1}{m_B+m_V} A_2(q^2) \,,\quad
a_-(q^2) = \frac{2m_V}{q^2}A_0(q^2) - \frac{m_B+m_V}{q^2} A_1(q^2) + 
\frac{m_B-m_V}{q^2} A_2(q^2)
\nonumber\\
g_+(q^2) &=& T_1(q^2)\,,\quad g_-(q^2) = 
\frac{m_B^2-m_V^2}{q^2}(T_2(q^2)-T_1(q^2))\,,\quad
h(q^2) = \frac{1}{q^2}(T_1(q^2)-T_2(q^2)) - 
\frac{1}{m_B^2-m_V^2} T_3(q^2)\nonumber\,.
\end{eqnarray}

In addition to these form factors, we require also the matrix elements of 
the dimension-4 operators $\bar q i\Dleft_\mu (\gamma_5) b$, which can be defined as 
\bea\label{Ddef}
\langle V(k,\eta) |\bar q i\Dleft_\mu b|\bar B(v)\rangle &=&
d(q^2) i\varepsilon_{\mu\nu\lambda\sigma} \eta^{*\nu} (p+k)^\lambda (p-k)^\sigma\\
\label{D1def}
\langle V(k,\eta) |\bar q i\Dleft_\mu \gamma_5 b|\bar B(v)\rangle &=&
d_1(q^2) \eta^*_\mu + d_+(q^2) (\eta^*\cdot p)(p_\mu + k_\mu)\\
&+& d_-(q^2) (\eta^*\cdot p)(p_\mu - k_\mu)\,.\nonumber
\eea
Their scaling with the heavy quark mass $m_b$ is complicated by the presence
of the covariant derivative $iD_\mu$, which can introduce factors of the large 
scale $m_b$ through
loops. To make it explicit, we consider the matching of the
dimension-4 QCD operators in Eqs.~(\ref{Ddef}), (\ref{D1def}) onto HQET. Working at
tree level in the dimension-4 operators, but keeping all contributions enhanced by
$O(m_b)$, this can be written as
\begin{eqnarray}\label{matchD1}
\bar q i\Dleft_\mu b &=& D_0^{(v)}(\mu) m_b \bar q \gamma_\mu h_v + 
D_1^{(v)}(\mu) m_b \bar q v_\mu h_v + \bar q i\Dleft_\mu h_v + \cdots\\
\label{matchD5}
\bar q i\Dleft_\mu \gamma_5 b &=& - D_0^{(v)}(\mu) m_b \bar q \gamma_\mu\gamma_5  h_v + 
D_1^{(v)}(\mu) m_b \bar q v_\mu\gamma_5  h_v + \bar q i\Dleft_\mu\gamma_5  h_v + \cdots
\end{eqnarray}
We assumed here the naive anticommuting $\gamma_5$ scheme. The Wilson coefficients
$D_i^{(v)}(\mu)$ start at $O(\alpha_s)$. 

The matrix elements of the dimension-4 HQET operators analogous to those appearing
in Eqs.~(\ref{Ddef}), (\ref{D1def}) (obtained by replacing $\bar q 
i\Dleft_\mu (\gamma_5) b \to \bar q i\Dleft_\mu (\gamma_5) h_v$)
can be parameterized in terms of similar form factors, 
denoted with $d^{(0)}(q^2), \dots$. They have a simple scaling with the heavy
quark mass, which is the same as in Eq.~(\ref{pc}) 
with the substitution $(d^{(0)},d_1^{(0)},d_+^{(0)},d_-^{(0)}) \to (g,f,a_+,a_-)$.
These form factors are related to the
effective theory form factors introduced in \cite{GrPi1} as
\bea
d^{(0)}(q^2) = \frac12 {\cal D}(q^2)  \,,\qquad 
d_1^{(0)}(q^2) = - {\cal D}_1(q^2)  \,,\qquad \cdots
\eea

Taking the $B\to V$ matrix element of Eq.~(\ref{matchD1}) one finds for the
leading terms in the $1/m_b$ expansion of $d(q^2)$ 
\begin{eqnarray}\label{dexp}
d(q^2) = \frac{D_0^{(v)}(\mu)}{C_0^{(v)}(\mu)} m_b g(q^2) + d^{(0)}(q^2,\mu) + \dots
\end{eqnarray}
We keep here all terms of order $O(\alpha_s m_b^{1/2})$ and $O(m_b^{-1/2})$
and the ellipses denote terms of order $O(\alpha_s m_b^{-1/2}, m_b^{-3/2})$.
Similar expansions are obtained from Eq.~(\ref{matchD5}) 
\begin{eqnarray}
d_1(q^2) &=& -\frac{D_0^{(v)}(\mu)}{C_0^{(v)}(\mu)} m_b f(q^2) + d_1^{(0)}(q^2) + \dots\\
d_+(q^2) &=& -\frac{D_0^{(v)}(\mu)}{C_0^{(v)}(\mu)} m_b a_+(q^2) + d_+^{(0)}(q^2) + \dots\\
d_-(q^2) &=& -\frac{D_0^{(v)}(\mu)}{C_0^{(v)}(\mu)} m_b a_-(q^2) + d_-^{(0)}(q^2) + \dots\,.
\end{eqnarray}

In the low recoil region, heavy quark symmetry predicts relations among
these form factors \cite{IsWi,BuDo}. The sub-leading corrections to these 
relations were computed in \cite{GrPi1}. We give here an alternative
simpler derivation, valid to all orders in $1/m_b$ (see also \cite{proc}). 
We take this
opportunity to include also $O(m_q)$ light quark mass effects 
(with $m_q$ the mass of the quark produced in the weak decay $b\to q$)
in these relations, which were neglected in \cite{GrPi1}.
Such effects can be important for the case of $B\to K^*$ decays.

The first relation is obtained from the operator identity
\begin{eqnarray}\label{ward1}
i\partial^\nu (\bar q i\sigma_{\mu\nu} b) = -(m_b+m_q) \bar q\gamma_\mu
b - 2\bar q i\Dleft_\mu b + i\partial_\mu(\bar q b)\,,
\end{eqnarray}
which follows from a simple application of the QCD equations of motion
for the quark fields. Taking the $B\to V$ matrix element one finds the
exact relation
\bea\label{W-1}
g_+(q^2) = -(m_b+m_q) g(q^2) -2 d(q^2)\,.
\eea
Counting powers of $m_b$ and keeping the leading order terms gives
the well-known Isgur-Wise relation among vector and tensor form factors
\cite{IsWi} $g_+(q^2) = -m_B g(q^2)$. Keeping also the subleading terms of 
$O(m_b^{-1/2})$ reproduces the
improved form factor relations derived in \cite{GrPi1}.
Inserting the expansion of $d(q^2)$ Eq.~(\ref{dexp}) into
Eq.~(\ref{W-1}) gives 
\begin{eqnarray}\label{W-1p}
g_+(q^2) = - \left(1 + 2\frac{D_0^{(v)}(\mu)}{C_0^{(v)}(\mu)}\right) m_b g(q^2) 
- 2 d^{(0)}(q^2) - m_q g(q^2)  + \cdots\,.
\end{eqnarray}
This agrees with the improved symmetry relation Eq.~(48) of Ref.~\cite{GrPi1} and 
generalizes it by including light quark mass effects and by making explicit the
renormalization scale dependence. The radiative corrections to this relation
were computed in Ref.~\cite{GrPi1} at $\mu=m_b$ in terms of a coefficient 
$\kappa_1$ (defined in Eq.~(23) of \cite{GrPi1}). Using Eq.~(\ref{D0exp})
below this coefficient can be expressed as
\begin{eqnarray}\label{kappa1}
\kappa_1(\mu) = \left(1 + 2\frac{D_0^{(v)}(\mu)}{C_0^{(v)}(\mu)}\right) 
\frac{m_b(\mu)}{m_B}
= \frac{C_0^{(t)}(\mu) - C_1^{(t)}(\mu)}{C_0^{(v)}(\mu)} \,.
\end{eqnarray}

The equation of motion Eq.~(\ref{ward1}) can be used to determine the 
Wilson coefficients $D_{0,1}^{(v)}(\mu)$ in the matching of the dimension-4 
operators Eq.~(\ref{matchD1}) in terms of the Wilson coefficients of the
dimension-3 currents. In this derivation we  set $i\partial_\mu = m_B
v_\mu - p_\mu = m_B v_\mu(1 + O(\Lambda/m_b))$. We find
\begin{eqnarray}\label{D0exp}
& &C_0^{(t)}(\mu) - C_1^{(t)}(\mu) = 
\frac{m_b(\mu)}{m_B} \left( C_0^{(v)}(\mu) + 2D_0^{(v)}(\mu) \right)\\
\label{D1exp}
& &C_0^{(t)}(\mu) - C_1^{(t)}(\mu) = -\frac{m_b(\mu)}{m_B} 
\left( C_1^{(v)}(\mu) + 2D_1^{(v)}(\mu) \right) + C_0^{(s)}(\mu)\,,
\end{eqnarray}
where $C_0^{(s)}(\mu)$ is the Wilson coefficient appearing in  the matching of the
scalar current in QCD onto HQET
\begin{eqnarray}\label{Cs}
\bar s b = C_0^{(s)}(\mu) \bar q h_v  + \cdots\,.
\end{eqnarray}
Another application of the equations of motion for the vector current
$i\partial^\mu (\bar q \gamma_\mu b) = (m_b-m_q) (\bar qb)$ determines this Wilson
coefficient in terms of those of the vector current as
\begin{eqnarray}
C_0^{(v)}(\mu) + C_1^{(v)}(\mu) = \frac{m_b(\mu)}{m_B} C_0^{(s)}(\mu)\,.
\end{eqnarray}
At the order we work, the B meson mass can be replaced with the $b$
quark pole mass, and the corresponding mass ratios in Eqs.~(\ref{D0exp}), 
(\ref{D1exp}) and (\ref{Cs}) are given by
\begin{eqnarray}
\frac{m_b(\mu)}{m_B} = 1 + \frac{\alpha_s C_F}{4\pi}
\big(-6\log\frac{\mu}{m_b} - 4\big)\,.
\end{eqnarray}
Combining these relations we find 
predictions for the Wilson coefficients $D_{0,1}^{(v)}(\mu)$, which are confirmed
also by explicit computation at one-loop order
\begin{eqnarray}
D_0^{(v)}(\mu) = \frac{\alpha_s C_F}{4\pi}\left( 2 \log\frac{\mu}{m_b} + 2\right)\,,
\qquad
D_1^{(v)}(\mu) = \frac{\alpha_s C_F}{4\pi}\left( 4 \log\frac{\mu}{m_b} + 2\right)\,.
\end{eqnarray}

The constraint Eq.~(\ref{D0exp}) can be used to relate the scaling of the
$1 + 2D_0^{(v)}(\mu)/C_0^{(v)}(\mu)$ factor to known anomalous dimensions.
It satisfies the RG equation
\begin{eqnarray}\label{gamD}
\mu \frac{d}{d\mu}\left(1 + 2\frac{D_0^{(v)}(\mu)}{C_0^{(v)}(\mu)}\right) = 
\gamma_D(\alpha_s) \left(1 + 2\frac{D_0^{(v)}(\mu)}{C_0^{(v)}(\mu)}\right)
\end{eqnarray}
with anomalous dimension $\gamma_D(\alpha_s) = - \gamma_t(\alpha_s) - 
\gamma_m(\alpha_s)$. We denoted here with $\gamma_t$ the anomalous dimension of the
tensor current defined as
\begin{eqnarray}\label{gamt}
\mu \frac{d}{d\mu} g_+(q^2) = -\gamma_t(\alpha_s) g_+(q^2) \,,\qquad
\gamma_t(\alpha_s) = \frac{2\alpha_s}{3\pi} + \cdots
\end{eqnarray}
and $\gamma_m$ is the mass anomalous dimension
\begin{eqnarray}\label{gamm}
\mu \frac{d}{d\mu} m(\mu) = \gamma_m(\alpha_s) m(\mu) \,,\qquad
\gamma_m(\alpha_s) = -\frac{2\alpha_s}{\pi} + \cdots\,.
\end{eqnarray}

Similar relations among the tensor and axial form factors 
are obtained starting with the operator identity 
(valid in the NDR anti-commuting $\gamma_5$ scheme)
\bea
i\partial^\nu (\bar q i\sigma_{\mu\nu}\gamma_5 b) = (m_b - m_q) \bar q\gamma_\mu\gamma_5
b - 2\bar qi\Dleft_\mu \gamma_5 b + i\partial_\mu(\bar q\gamma_5 b)\,.
\eea
Taking the $B\to V$ matrix element gives three relations
\begin{eqnarray}\label{W-2}
& &(m_B^2-m_V^2) g_+(q^2) + q^2 g_-(q^2) = (m_b-m_q) f(q^2) - 2d_1(q^2)\\
\label{W-3}
& &- g_+(q^2) + q^2 h(q^2) = (m_b - m_q) a_+(q^2) - 2 d_+(q^2)\\
& & - g_-(q^2) - (m_B^2-m_V^2) h(q^2) = (m_b - m_q) a_-(q^2) - 
2d_-(q^2) + s(q^2)
\end{eqnarray}
After using here the $1/m_b$ expansions for the $d_{1,+,-}(q^2)$ form factors,
we find the final form of the symmetry relations to subleading order in $1/m_b$
\begin{eqnarray}\label{W-2p}
& &(m_B^2-m_V^2) g_+(q^2) + q^2 g_-(q^2) = (1+2D_0^{(v)}(\mu)/C_0^{(v)}(\mu))(m_b-m_q) f(q^2) - 
2d_1^{(0)}(q^2) + \cdots\\
\label{W-3p}
& &- g_+(q^2) + q^2 h(q^2) = (1+2D_0^{(v)}(\mu)/C_0^{(v)}(\mu)) (m_b - m_q) a_+(q^2) - 
2 d_+^{(0)}(q^2) + \cdots\\
& & - g_-(q^2) - (m_B^2-m_V^2) h(q^2) = (1+2D_0^{(v)}(\mu)/C_0^{(v)}(\mu)) (m_b - m_q) a_-(q^2) - 
2d_-^{(0)}(q^2) + \cdots
\end{eqnarray}
Together with Eq.~(\ref{W-1p}), these relations are of phenomenological significance 
and are used in Sec.~V to express the contribution of the electromagnetic
penguin $Q_7$ to the $B\to K^* \ell^+\ell^-$ amplitude.

We illustrate in the following the application of Eq.~(\ref{W-2}) to give
an alternative derivation of the power correction to a heavy quark symmetry 
relation presented in \cite{GrPi1}.
Consider the combination of form factors
\bea\label{calF}
{\cal F}(q^2) &=& (m_B + m_V) g_+(q^2) + (m_B-m_V) g_-(q^2) \,.
\eea
The relation Eq.~(\ref{W-2}) gives a prediction for ${\cal F}(q^2) $
at the zero recoil point $q^2_{\rm max}=(m_B-m_V)^2$
\bea
{\cal F}(q^2_{\rm max}) =
\left( 1 + \frac{m_V - \bar\Lambda - m_q}{m_B}\right) f(q^2_{\rm max}) -
\frac{2}{m_B} d_1^{(0)}(q^2_{\rm max})\,.
\eea
The leading term on the right-hand side was obtained in \cite{IsWi,SaYa} and the 
$1/m_b$ correction was given in \cite{GrPi1} (we correct here the sign of the 
$O(1/m_b)$ term in the brackets).

\end{document}